\documentclass[fleqn,usenatbib]{mnras}

\usepackage{newtxtext,newtxmath}

\usepackage[T1]{fontenc}
\usepackage{ae,aecompl}
\usepackage{lscape}

\usepackage{graphicx}	
\usepackage{amsmath}	
\usepackage{amssymb}	
\usepackage{nccmath}    
\usepackage{multirow} 
\usepackage{multicol} 
\usepackage{booktabs} 
\usepackage{float} 

\title[Transfer learning for radio galaxy classification]{Transfer learning for radio galaxy classification}

\author[H. Tang et al.]{
H. Tang,$^{1}$\thanks{E-mail: hongming.tang@postgrad.manchester.ac.uk (HT)}
A. M. M. Scaife,$^{1}$
and J. P. Leahy$^{1}$
\\
$^{1}$Jodrell Bank Centre for Astrophysics, University of Manchester, Oxford Road, Manchester M13 9PL, UK
}

\date{Accepted XXX. Received YYY; in original form ZZZ}

\pubyear{2018}

\hypersetup{draft}
\begin{document}
\label{firstpage}
\pagerange{\pageref{firstpage}--\pageref{lastpage}}
\maketitle

\begin{abstract}
In the context of radio galaxy classification, most state-of-the-art neural network algorithms have been focused on single survey data. The question of whether these trained algorithms have cross-survey identification ability or can be adapted to develop classification networks for future surveys is still unclear. One possible solution to address this issue is $transfer~learning$, which re-uses elements of existing machine learning models for different applications. Here we present radio galaxy classification based on a 13-layer Deep Convolutional Neural Network (DCNN) using transfer learning methods between different radio surveys. We find that our machine learning models trained from a random initialization achieve accuracies comparable to those found elsewhere in the literature. When using transfer learning methods, we find that inheriting model weights pre-trained on FIRST images can boost model performance when re-training on lower resolution NVSS data, but that inheriting pre-trained model weights from NVSS and re-training on FIRST data impairs the performance of the classifier. We consider the implication of these results in the context of future radio surveys planned for next-generation radio telescopes such as ASKAP, MeerKAT, and SKA1-MID.


\end{abstract}

\begin{keywords}
radio continuum: galaxies -- methods: statistical -- surveys
\end{keywords}



\section{Introduction}
\label{sec0_0}






The radio morphology of Double Radio sources associated with Active Galactic Nuclei \citep[DRAGNs;][]{1993LNP...421....1L} is typically determined by examining the distribution of high and low surface brightness regions in the radio synchrotron emitting relativistic jets associated with these systems. This morphology
is the basis of the well known Fanaroff-Riley radio galaxy classification scheme \citep[FR;][]{1974MNRAS.167P..31F}. This classification splits radio galaxies into two groups, known as FR~I and FR~II, respectively. Sources where the ratio of the distance between the highest brightness regions (known as hotspots) to the total extent of the source is smaller than 0.5.
are classified as FR~I. Radio galaxies with larger distance ratios are identified as FR~II sources \citep{1974MNRAS.167P..31F}. The FR classification scheme is also known to be related to a number of other morphological characteristics \citep{1993LNP...421....1L} including jet opening angle \citep{1984ARA&A..22..319B,2014MNRAS.441.1488P}. Moreover, although FR classification is primarily made on source morphology it is also strongly correlated with radio luminosity \citep{1974MNRAS.167P..31F}.



Traditionally, identifying FR class is done by visual inspection and 
such an approach has been widely used over the past few decades \citep{1996MNRAS.279..257S,1996MNRAS.281.1081C,1996IAUS..175..157L,1999MNRAS.309..100I,2001AA...374..861S,2001AA...370..409L,2001AA...371..445M,2005AJ....130..896S,2006AA...454...85M,2006AA...454...95M,2007AcA....57..227M,2011AstBu..66..416S,2012MNRAS.426..851K,2016ARep...60..718B,2017MNRAS.469.2886D}. However, we note that this approach is constrained by the resolution and sensitivity of given radio data. With poor resolution, FR~IIs with extended structure may be only partially resolved, with no clear hotspots, and thus be misclassified as FR~Is

A visual inspection method has been widely used in part due to radio survey sample sizes. For instance, the NRAO VLA Sky Survey \citep[NVSS;][]{1998AJ....115.1693C}, Sydney University Molonglo Sky Survey \citep[SUMSS;][]{1999AJ....117.1578B}, and the Faint Images of the Radio Sky at Twenty-Centimeters \citep[FIRST;][]{1995ApJ...450..559B} have cataloged sources of no more than 2 million objects, respectively \citep{1997ApJ...475..479W,1998AJ....115.1693C,2003MNRAS.342.1117M}. 
Although still large, the size of these catalogues enabled the possibility of finishing the work above by visual inspection. Such sample sizes also make citizen science projects like Radio Galaxy Zoo \citep[RGZ;][]{2015MNRAS.453.2326B} possible. 

Radio catalogs produced by the next generation of radio surveys, however, are anticipated to be much larger. The Australia SKA Pathfinder \citep[ASKAP;][]{2008ExA....22..151J} Evolutionary Map of the Universe \citep[EMU;][]{2011PASA...28..215N} survey is expected to produce a catalog of $\sim$70 million sources. Among the objects in this catalog, around 7 million extended radio sources are likely to require visual inspection. 

%
Beyond ASKAP, the Square Kilometre Array \citep[SKA;][]{2004NewAR..48..979C} will be able to observe FR~Is, low- and high-luminosity FR~IIs up to redshift z$\sim$4 \citep{2015aska.confE.173K}. This telescope array will have better resolution and sensitivity than current best radio telescopes \citep{2015aska.confE..83M} and thus should also discover a huge number of extended radio sources.

The anticipated volume of objects from new radio surveys have motivated the introduction of semi-automatic and automatic object classification algorithms. Recently, convolutional neural networks \citep[CNNs;][]{Krizhevsky2012} were applied to radio galaxy morphology classification \citep{2017ApJS..230...20A,2018arXiv181207190M}. These studies suggest that complex radio source structures can be identified and classified according to their morphology from images drawn from a single survey. 
%
%
However, although such machine learning approaches are useful, training a model often requires a considerable amount of labeled data. One practical way to collect these labels is through citizen science. Radio Galaxy Zoo, a representative astronomical citizen science project, is primarily used to cross-identify infrared hosts and their corresponding radio lobes. The project has $\sim$12,000 users and has already made over 2 million classifications since it launched. These classifications have contributed to the foundation of the RGZ DR1 dataset \citep[RGZ Data Release 1;][]{Wong2018}. The dataset contains over 70,000 candidate sources, with source coordinates, fluxes, angular extent recorded. The project team have now developed several machine learning algorithms for doing infrared-radio source cross-identification, compact-resolved source separation, radio galaxy localization and peak-component morphology classification with the dataset \citep{2018MNRAS.476..246L,2018arXiv180512008W,2018MNRAS.478.5547A}. 

Although citizen science can contribute to dataset foundation, collecting data can take multiple years. In addition, it is still unclear whether such an approach is transferable to other survey data. How to build a dataset cost-efficiently and maximize the generalisation of machine learning models across multiple surveys remain open questions. One solution to these two questions, however, might be transfer learning \citep{2014arXiv1411.1792Y} 


In this paper, we investigate the applicability of transfer learning to the classification of radio galaxy morphology using survey data. In \S~\ref{sec1_1} we introduce the use of CNNs for classification and in \S~\ref{sec1_2} we describe the practice of transfer learning. Construction of our training, test and validation datasets, including data acquisition, image pre-processing and data formatting, is described in \S~\ref{sec2_0}. \S~\ref{sec3_0} covers the network architecture adopted in this study and the transfer learning strategies. In \S~\ref{sec4_0}, we compare and discuss the performance of these strategies, and in \S~\ref{sec5_0} we discuss the applicability of our results to future radio surveys.

In what follows we assume a $\rm \Lambda$CDM cosmology with $\Omega_{\rm m}$ = 0.3153, $\Omega_{\Lambda}$ = 0.6847, and a Hubble constant of $\rm H_{\rm 0} = 67.36~km~s^{-1}~Mpc^{-1}$ \citep{2018arXiv180706205P}. Computational work in this paper was done using a system with 32 Intel Xeon(R) E5-2640 v3 CPUs at 2.6\,GHz and 202.5\,GB memory.

\section{Theory}
\label{sec1_0}

\subsection{Classification using Neural Networks}
\label{sec1_1}

In recent years, convolutional neural networks (CNNs) have been widely adopted for a variety of applications in image recognition  \citep[e.g.][]{2014arXiv1409.0575R}, video analysis \citep[e.g.][]{Huang2018} and natural language processing \citep[e.g.][]{Grefenstette2014}. For astronomy, CNNs have been used to identify galaxy clusters and filaments \citep{2018MNRAS.480.3749G}, detect fast radio bursts \citep{2018arXiv180303084C}, localize radio galaxies and identify their morphologies \citep{2017ApJS..230...20A,2018arXiv180512008W}, recognize strong gravitational lenses \citep{2018arXiv180203609M}, and to classify supernovae \citep{2017arXiv1711.11526K}. 

CNNs are widely accepted in image classification for several reasons. Firstly, CNNs decompose images into multiple patches that are partially overlapped, where each cortex neuron only corresponds to a single patch \citep{Matsugu2003}. This enables the network to classify images with little data pre-processing or a priori feature value designation \citep{Krizhevsky2012}. Secondly, CNNs are \textit{weight-sharing}. With shared weights, CNNs can learn features from a single sample via back-propagation and then share those weights with other samples. Thirdly, as a consequence of weight-sharing, convolutional networks are translationally invariant. This last characteristic leads to efficiency as it decreases the required input sample size necessary to train a model with good prediction capability.

CNNs use a number of multi-layer perceptrons \citep[MLP;][]{Rosenblatt1961} to learn both machine learning features and classification weights through a supervised learning method known as back-propagation \citep{Goodfellow2016}. Apart from the first and the final layer of a network, every layer treats outputs from the previous layer as inputs and forwards its own outputs to the following layer. Inputs to the first layer are sample images, potentially with multiple channels (e.g. 1 for a grayscale image or 3 for an RGB image). Outputs from the final layer are used for class prediction. The layers between the input and output layer are known as \textit{hidden} layers and include convolutional layers, fully connected layers, pooling layers, and loss layers. The functionality of these layers will be described below.

For CNNs, multi-layer perceptrons are created by applying non-linear activation functions to every convolutional and fully-connected layer output. These activation functions introduce non-linearity to the network \citep{Duda2012}. When a network is linear, multi-layer perceptrons can be summed up to perform as a single-layer perceptron. A single-layer perceptron produces an output that is a linear combination of the inputs and thus has limited ability for feature recognition and classification. The addition of non-linearity at each layer boosts the expressive ability of a network \citep{Minsky1988}. Representative activation functions include the sigmoid activation function and the Rectified Linear Unit \citep[ReLU;][]{Hahnloser2000} function. Defined as {\tt f(y) = max(0,y)}, the ReLU function can help to build sparser models and is therefore beneficial when building deeper training networks \citep{Glorot2011}. \cite{2017arXiv170206257C} claim that, compared to the sigmoid function, models trained with the ReLU activation function have higher efficiency and better overall precision.

CNNs learn via a method called ``backward propagation of errors" \citep[][]{Goodfellow2016} combined with gradient descent optimization algorithms \citep{Sebastian2016}. The goal of this learning is to minimize the loss function between observation and prediction \citep{Song2016, Heas2018}. In the context of  categorical classification, a representative loss function called {\it categorical cross entropy}, {\tt H(y,q)}, is generally used:

\begin{ceqn}
\begin{align}
\rm H({\tt y},{\tt q}) = -({\tt y}~ln({\tt q}) + (1-{\tt y})~ln(1-{\tt q})).
\label{eqn:X}
\end{align}
\end{ceqn}

For FR binary classification, {\tt y} indicates if class label (0 or 1) is classified as FR~I for an observation. {\tt q}, on the other hand, refers to the predicted probability that an observation is of FR~I class. Ideally, if the predictions match the observations perfectly, {\tt H(y,q)} will become zero.  

Overall, CNN optimization can be summarized as a two-phase circular process of propagation and weight updates. During the propagation phase, input samples propagate through the network and form a network output based on initial network weights and biases. The network then compares the desired model output and the actual network output by measuring their cross entropy. Calculated errors for the output layer then propagate back through the network. Backpropagation finishes when all neurons in the network have equivalent error values. The backpropagation method then uses these values to calculate the gradient of the cross entropy loss. The resulting gradient becomes the starting point of the optimization phase. In this second phase, backpropagation passes the gradient to a gradient descent optimization algorithm and updates the network weights.

\vskip .1in
\noindent
In practice, a typical convolutional network contains:
\begin{itemize}
	\item \textbf{Convolutional layers:} Each convolutional layer uses a number of learnable filters. The specified number of filters is known as the \textit{depth} of the layer. Each filter is applied across the full volume of the input data, iteratively calculating the local dot product of the filter with the input. Often a convolutional layer will be combined with an activation step, which re-introduces non-linearity into the network following the linear convolution operation. A typical activation function that is applied to the output of a convolutional layer is the ReLU function described above.
	\item \textbf{Pooling layers:} Pooling layers provide a form of down-sampling. A pooling layer partitions the input from the previous layer into non-overlapping segments and calculates a spatial pooling function for that segment. The function applied is designed to preserve the most important information in each case. Such functions may include returning the maximal value for that segment (max-pooling) or the average value (average-pooling). Pooling layers reduce the dimensionality of the data within the network and hence the required computation; they also have the effect of reducing over-fitting by making the network insensitive to small distortions and variations in the data. 
    \item \textbf{Fully Connected layers:} Fully connected layers are used to implement classification within the network. Neurons within fully connected layers are connected to all activated outputs in the previous layer. Activations can then compute with learnable filters by matrix computation and a bias offset.
    \item \textbf{Loss layer:} The loss layer usually acts as the final network layer. Its outcome often represents a categorical distribution and evaluates how the network prediction deviates from the truth.
\end{itemize}

With 2 convolutional layers, 2 max-pooling layers, and 1 fully-connected layer, \cite{2018MNRAS.480.3749G} were able to generate a classifier to identify candidate galaxy clusters and filament structures from simulated image data. The model correctly classified more than 90\% of the unlabelled test images, i.e. the model accuracy was higher than 90\%.

In addition to the standard network layers described above, dropout layers \citep{Srivastava2014} and batch normalization layers \citep{2015arXiv150203167I} have also been developed. The purpose of the dropout function is to remove weakly connected neurons and thus avoid over-fitting. Batch normalization (BN), on the other hand, is designed to speed up the training process and regularize a neural network. If each learnable layer of a model uses either batch normalization or dropout functions, the model can become more computationally efficient especially in the context of image classification \citep{Szegedy2015,2017arXiv170404861H}. Meanwhile, a models which simultaneously applied both functions on each learnable layers reported impaired or poor performance \citep{2015arXiv150203167I,2018arXiv180105134L}. It seems that an appropriate strategy is to choose one out of the two functions for each learnable model layer.

A representative example is a network designed by \cite{Krizhevsky2012}. The 12-layer convolutional network applies BN on convolutional layers and Dropout on fully-connected layers. In the context of an astronomical applications, \cite{2017ApJS..230...20A} slightly modified this architecture, and their model reached over 90\% accuracy when performing three-class radio galaxy morphology (FR~I, FR~II, and Bent-tailed sources) classification with given test images.

\subsection{Transfer learning}
\label{sec1_2}
Constrained by the lack of large-scale labelled astronomical training data, as well as limited computational power, recent astronomical applications have turned their attention to pre-trained models \citep{2018arXiv180512008W}. These models were trained with ImageNet, a sizeable visual dataset with a thousand object categories and millions of labeled sample images from daily life \citep{2014arXiv1409.0575R}. As expected, the performance of these generalized models when applied to astronomical survey data was found to be inferior to that of custom models for direct classification \citep{2018MNRAS.476..246L}.


An alternative to direct classification using inherited weights is to initialize from pre-trained model weights before performing customized training. One can restore model weights for either just the initial \citep{2018arXiv180512008W} or all layers in a network \citep{2018MNRAS.479..415A}. In some cases, such an approach has been shown to help customized models to avoid over-fitting and/or to help them to accelerate the training process \citep{2018MNRAS.479..415A,2018arXiv180512008W}. 

Known as {\it transfer learning}, this approach can re-use knowledge from a solved problem and apply it to relevant but different applications \citep{Pratt1993}. For instance, learning general features from handwritten digits and applying them in handwritten character recognition \citep{Maitra2015}. In the context of classifying FR morphology in radio galaxies with CNNs, generic features learned from initial network layers, such as source edges and hot spots, should exist irrespective of the radio survey; complex features, however, would be learned by the last few layers \citep{2017ApJS..230...20A}. Using this approach to mitigate against model over-fitting during the training process is referred to as regularization \citep{2018MNRAS.479..415A}.


The application of transfer learning for classification requires careful consideration of which layers to train \citep{2017arXiv170901476S}. When applying transfer learning, a network is trained on a new dataset of samples using the stored weights for some or all layers from a pre-trained network as an initialization, rather than initializing weights randomly. The choice of which layers to restore depends on the size of the dataset used to train the pre-trained model, the size of the new dataset, the correspondence of the two datasets, and the architecture of the pre-trained network. 

Here, we consider the use of pre-trained models trained on radio survey data. Rather than learning everyday object features, layers of the CNN learn radio galaxy features such as jets and hot spot relative positions. These features are universal in classifying source FR morphology, irrespective of survey.

The application of transfer learning has two major advantages. Firstly, by inheriting general morphological features from a pre-trained model, a new model may have a better starting point. Secondly, freezing the weights for the convolutional layers can  significantly reduce the training time required to achieve comparable accuracy. However, although loading weights from pre-trained models is often practical, the method needs to be treated carefully. High-level features needed to be trained on customized datasets to learn features for a specific classification problem. In addition, as a pre-trained model and customized model are usually addressing different objectives, transfer learning might produce a {\tt NaN} loss function value \citep{2018arXiv180512008W}. 

Furthermore, transfer learning strategies vary depending on network architecture and dataset definitions \citep{2017ApJS..230...20A}. Inappropriate architectures can lead to over-fitting, which can become severe when the re-trained model uses only a small number of new training samples but has a considerable number of learnable parameters \citep{2018arXiv180512008W}. The relative training sample size for pre-trained and transfer-learning models requires careful consideration in order to obtain good model performance. 

In the context of classifying radio morphology, the influence of transfer learning using data from surveys at different frequencies and with different resolution remains unclear. Here, we explore this question using a transfer learning approach implemented on a variation of the AlexNet CNN \citep{Krizhevsky2012,2017ApJS..230...20A}. As part of this work:


\begin{itemize}
	\item \textbf{} We develop a quasi-automatic pipeline to construct training datasets from archival radio surveys. This can be used to download and process images from various surveys. This pipeline  makes comparing models trained on different surveys possible.
	\item \textbf{} We convert the data samples to a dataset format consistent with the standard MNIST machine learning dataset \citep{Lecun1998}. This enables the datasets to be used in other network architectures.
    \item \textbf{} We simplify and modify the AlexNet CNN architecture \citep{Krizhevsky2012}, a widely accepted CNN architecture recently adopted in radio galaxy classification \citep{2017ApJS..230...20A}. The primary network requires parallel GPU computation, which is not considered in this work. Our resulting network can be trained and tested with modest computation power to provide end-to-end training and classification.
    \item \textbf{} We develop and implement three different transfer learning strategies. Pre-trained models are trained on either NVSS \citep{1998AJ....115.1693C} or FIRST \citep{1995ApJ...450..559B} images, with final transfer learning models  transfer-learned on the other.  
    \item \textbf{} We demonstrate that the architecture can be used to classify FR radio morphology and achieve accuracy comparable with the performance of human radio astronomers.
	\item \textbf{} We evaluate the feasibility, training time cost, and model performance when applying different transfer learning strategies. We examine the possibility of using the same classifier to make prediction on both NVSS and FIRST images. 
\end{itemize}

This work provides an alternative method to full network training for radio galaxy classification and demonstrates under what circumstances such an approach is valid. 

\section{Data Sample Construction}
\label{sec2_0}
The data sample used in this work is designed to train a machine learning model capable of automatically classifying FR~I and FR~II radio galaxies. We use input data extracted from the NVSS \citep{1998AJ....115.1693C} and FIRST \citep{1995ApJ...450..559B} radio surveys. Here we describe the data selection process as well as the automated acquisition methods used.

\subsection{Astroquery based image data batch download}
\label{sec2_1}





We selected an input sample of radio galaxies following a similar methodology to \cite{2017ApJS..230...20A}. This method uses
the Combined NVSS and FIRST Galaxies catalogs \citep[CoNFIG;][]{2008MNRAS.390..819G,2010MNRAS.404.1719G}, and the FRICAT catalog \citep{2017A&A...598A..49C} to select objects from the FIRST surveys. These catalogs were selected as they share significant source populations, the data are free accessible, and sources are well-resolved in the FIRST images \cite{2017ApJS..230...20A}.

The CoNFIG catalog contains 859 resolved sources divided into 4 subsamples (CoNFIG 1, 2, 3 \& 4) with flux density limits of $ S_{\rm 1.4\,GHz} \ge$ 1.3, 0.8, 0.2, 0.05\,Jy, respectively. These sources are selected from the NVSS survey within the northern field of the FIRST survey.


Redshifts for the catalog sources were obtained either from SIMBAD, the NASA Extragalactic Database (NED) or estimated by using the $K_{s} - z$ relationship \citep{2010MNRAS.404.1719G} with source $K_{s}$  magnitudes from the 2MASS survey \citep{2006AJ....131.1163S} where available. 
From the full sample, 638 sources were associated with redshifts.

Source morphologies in the CoNFIG catalog were manually identified by looking at their NVSS and FIRST contour maps. Objects were classified as FR~I, FR~II, \textit{compact}, or \textit{uncertain} \citep{2010MNRAS.404.1719G}. Sources with collimated jets, showing hotspots close to their cores, were classified as FR~I; sources with their lobes aligned and hotspots situated at the edges of the lobes were classified as FR~II. If the source morphology was ambiguous, it was classified as \textit{uncertain}. 
 Sources with sizes smaller than 1$''$ were classified as \textit{compact}. In all, 95.7$\rm \%$ of CoNFIG sources have their radio morphology classified. We note that the FR~I sample in this catalog also includes sources that are considered to be wide-angle tail or irregular by other studies \citep{1993LNP...421....1L}. 

Depending on whether a candidate in the CoNFIG sample showed a clear FR~I or FR~II morphology, the identification of each object was qualified as `confirmed' (c) or `possible' (p). `Confirmed' sources were confirmed using the VLBA Calibrator Surveys \cite[VCS;][]{2002ApJS..141...13B,2003AJ....126.2562F,2006AJ....131.1872P,2007AJ....133.1236K} or the Pearson-Readhead survey \citep{1988ApJ...328..114P}. The final catalog contains 50 confirmed FR~I objects and, 390 confirmed FR~II sources.


To balance FR~I and FR~II sample sizes, the FRICAT catalog of FR~I radio galaxies was introduced . This catalog, which contains 219 FR~I radio sources, is a subsample of the catalog from \cite{2012MNRAS.421.1569B}, hereafter BH12. BH12 was formed by cross-matching the optical spectroscopy catalogs produced by \cite{2004MNRAS.351.1151B,2004ApJ...613..898T} based on data release 7 of Sloan Digital Sky Survey \citep[SDSS DR7;][]{2009ApJS..182..543A} with the NVSS and FIRST surveys, for sources with flux densities in NVSS that were greater than 5\,mJy.

To make the FRICAT sample, the authors compiled all BH12 sources with $z<0.15$, resulting in 3,357 objects. From this sample, all sources with radio emission extended more than 30\,kpc from the host galaxy center were selected, resulting in a sample of 741 sources with well-resolved morphologies. From these 741 sources, individual objects were then classified as FR~I type if (1) a one-sided or two-sided jet was present (including sources with bent jets), (2) the surface brightness of the jet decreased along its length, and (3) there was little or no brightness enhancement at the end of jet. 

Sources in which brightening was observed along the jet \citep[e.g. wide-angle tailed sources][]{1982IAUS...97...45B}, were excluded. All three FRICAT authors classified the sources independently, and an object would be added to the catalog only if 2 out of 3 authors agreed. The final catalog contains 219 FR~I radio galaxy candidates. The hosts of the FRICAT sources were found to be luminous ($-24 \ge M_{r} \ge -21$) early-type galaxies with black hole masses in the range $10^{8}$ to $3 \times 10^{9}$\,M$_{\odot}$.


CoNFIG and FRICAT together form the input sample for this work. By cross-matching FR~I source centroid coordinates from FRICAT and CoNFIG, we found 3 duplicate sources. To remain consistent with \cite{2017ApJS..230...20A}, we did not remove this small number of duplicate objects. The final sample comprises 266 FR~I sources and 390 FR~II sources. We extracted the central coordinates of host in each case and used these to download FIRST images for each object using the python {\tt astroquery} and {\tt urllib} tools. 

The {\tt astroquery} library is an affiliated package of {\tt astropy}, with tools for querying astronomical web forms and databases. We used the {\tt astroquery.skyview.get\_image\_list} function, specifying central source coordinates, setting survey name to {\tt $'$VLA~FIRST~(1.4~GHz)$'$} or {\tt $'$NVSS$'$}, and specifying the image scaling to be {\tt$'$Linear$'$}. With the returned list of image URLs we then used the {\tt urllib.request.urlretrieve} function to download NVSS and FIRST images. {\tt SkyView} by default returns each FITS image with size of 300 $\rm \times$ 300 pixels ($0.15 \rm \times 0.15$\,degrees for FIRST images). The pixel values in these  FITS images correspond to a brightness scale in units of Jy/beam.



\subsection{Image pre-processing and augmentation steps}
\label{sec2_2}



\begin{figure}
\setlength{\unitlength}{1cm}
\begin{center}
\includegraphics[width=9cm]{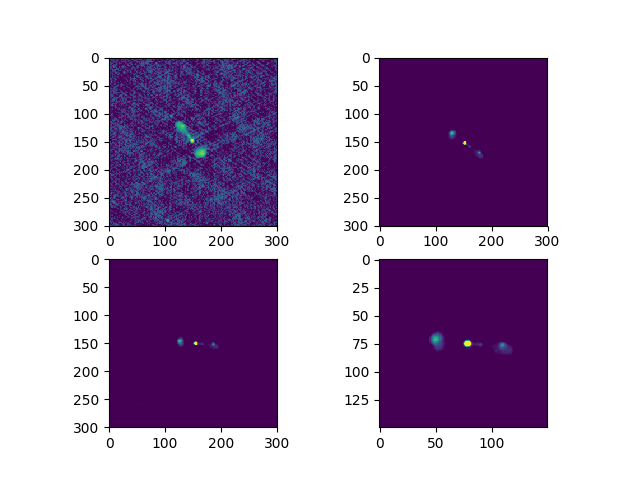}
\end{center}
\caption{An example of image pre-processing and augmentation. The upper left image is the log scaled original image downloaded from SkyView. The other three images, from left to right, top to bottom are the ones experienced sigma-clipping, rotation, and centered crop. The radio source centered at the sample FIRST image is 4C 31.30, a `confirmed' CoNFIG FR~II sample. The radio galaxy host locates at (J2000) 07:45:42.13 +31:42:52.6.}
\label{fig:pre_processing}
\end{figure}


The image pre-processing in this work consists of three operations: pixel-value re-scaling, image rotation and image clipping. These are performed on all input images.

\cite{2017ApJS..230...20A} reported that image background noise decreased classifier performance. Having investigated various noise clipping options, they proposed a solution where  pixel values lower than 3 times the local rms noise were set to zero, which we followed.

After clipping, we re-scaled each image following:
\begin{ceqn}
\begin{align}
\rm Output = \frac{Input - Min}{Max - Min} \times (255.0 - 0.0),
\label{eqn:I}
\end{align}
\end{ceqn}
where {\tt Max} and {\tt Min} refer to the maximum and minimum pixel value in an image. {\tt Output} and {\tt Input} represent the final re-scaled and original pixel values in the image, respectively. Re-scaled images have pixel values in the range from 0 to 255. These are then saved in PNG format. 

When training machine learning models, training datasets typically have sizes of order $\sim$10,000 data samples \citep{2017ApJS..230...20A}. 
Here, the original dataset contains a total of 659 source images, where $\sim$30$\%$ will serve as test samples to evaluate our trained model performance \citep{2017ApJS..230...20A}. Data augmentation was therefore considered to be necessary. 

Considering both dataset sample size and class balance, we decided to augment our dataset to have around 14,000 training images and 6,000 testing images for each class. Following previous practice \citep{2017ApJS..230...20A},
data augmentation was done by rotation of the original input source images by $1^{\circ}$, $2^{\circ}$, $3^{\circ}$ etc. Table~\ref{tab:Data_Augmentation} provides details of our dataset sample size. The final dataset contains 39,796 sample images.

\begin{table}

\begin{center}
\begin{tabular}{| c | c | c | c |}
\hline \hline
\multirow{ 2}{*}{\textbf{Class}} & \multicolumn{3}{ c |}{\textbf{Training/Validation}}   \\ 
\cline{2-4}
& \multicolumn{2}{ c |}{\textbf{Original}}   & \textbf{Augmented}    \\
\hline
FR~I & \multicolumn{2}{ c |}{189}   & 13,797   \\ 
FR~II & \multicolumn{2}{ c |}{273}   & 13,650  \\ \hline 
Total & \multicolumn{2}{ c |}{462}   & 27,447  \\ \hline
\multirow{ 2}{*}{\textbf{Class}} & \multicolumn{3}{ c |}{\textbf{Test}}   \\ 
\cline{2-4}
& \multicolumn{2}{ c |}{\textbf{Original}} & \textbf{Augmented}    \\
\hline
FR~I & \multicolumn{2}{ c |}{80} & 5,840   \\ 
FR~II & \multicolumn{2}{ c |}{117}  & 5,850  \\ \hline 
Total & \multicolumn{2}{ c |}{197}  & 11,690  \\ \hline \hline
\end{tabular}
\end{center}
\caption{A summary of FR~I, FR~II images of the dataset samples. The dataset consists of samples for training, validation, and testing. The augmented samples are created by the process claimed in Figure \ref{fig:pre_processing}. In step of $1^{\circ}$, FR~I source images were rotated from $1^{\circ}$ to $73^{\circ}$.  For FR~II images, we rotated them from $1^{\circ}$ to $50^{\circ}$. }
\label{tab:Data_Augmentation}
\end{table}

The final step in building our dataset was to clip the image sizes. Image clipping is to designed to constrain an image to its central source, yet remain large enough to identify structure. 
\cite{2017ApJS..230...20A} clipped sample images from their centers to 150 $\times$ 150 pixels, which is equivalent to a physical extent of 274.1\,kpc at $z=0.05$ for FIRST images. For NVSS image at the same redshift, however, image physical extent equals to 2283.8\,kpc. For the CoNFIG and FRICAT samples, 96.4$\%$ of them have their redshifts $\le$ 0.05. We visually inspected sample images from our dataset clipped in the same way, and found that the image width was suitable to recognize the sources while retaining characteristics sufficient to identify their FR morphology.   

\subsection{Data formatting and division}
\label{sec2_3}

After pre-processing, we converted our input dataset images into 
{\tt numpy} format. Target classification labels were defined as one-hot vectors: FR~I samples are labeled as [1., 0.], while FR~II samples have the label [0., 1.]. Section~\ref{sec3_0} discusses why such label vectors are convenient to use and will simplify the computational process when training the model. Label and image datasets were saved in 2-dimensional arrays. Rather than saving images in 3-dimensional arrays, we saved data input in 4-dimensional arrays as they are more flexible, capable of saving both single channel images (greyscale) and multiple channel images (e.g., RGB), and share the same format as the well-known MNIST dataset \citep{Lecun1998}. We also note that data were fully shuffled before being imported into the CNN.


We split the master dataset into training, validation, and test subsets. Each subset of data includes both image data and target classification labels. The training set is used to train the CNN machine learning model via back propagation. The validation set, on the other hand, helps us to examine whether the model is over-fitting the data through forward propagation. This examination takes place after every epoch training epoch. Validation subset samples can therefore be extracted from the original training set. The test data subset is separated from the training set and validation set, with samples unseen by model before testing. This subset provides samples for evaluating the performance of the trained model when doing realistic classification. Table~\ref{tab:Data_Augmentation} gives an overview of the data subsets used in this work. 

In the work, consistent with \cite{2017ApJS..230...20A}, we separated training and test samples with a ratio of 70-30. We then pre-processed and applied image augmentation on all sample images. The primary training set is split into training and validation using a ratio of 80:20.

\section{Network Architecture}
\label{sec3_0}

An appropriate network architecture should consider object complexity and computational power. If necessary, transfer learning ability is a factor as well. Although simple networks may perform well for some applications \citep[e.g.][]{2018MNRAS.480.3749G}, classifying radio galaxy morphology requires comparatively deeper networks. Early attempts have shown that, for radio galaxy classification, simple networks perform only slightly better than random guesses \citep{2017ApJS..230...20A}. With fewer learn-able layers, simple networks may also meet issues when using transfer learning as they have weaker expressive ability \citep{Oquab2014}. A reasonably deep network therefore becomes necessary for classifying radio galaxies and being capable of doing so using transfer learning. 

Among the pre-existing architectures, the AlexNet CNN is a representative deep network \citep{Krizhevsky2012}. This is a widely used 12-layer parallel computing CNN with 5 convolutional layers, 3 max pooling layers, 3 fully-connected layers, and a softmax readout layer. 
\cite{2017ApJS..230...20A} slightly modified this network and succeeded in classifying FR~I, FR~II, and Bent-tailed radio sources, with a general accuracy of above 90\%. Their network adopted a GPU-based implementation and was able to train 30 epochs of data in $\sim$70 mins \citep{2017ApJS..230...20A}. The number of epochs in machine learning refers to the number of times that the complete training dataset is imported during training. 

\begin{figure}
\setlength{\unitlength}{1cm}
\begin{center}
\includegraphics[width=8cm,height=0.9\textheight]{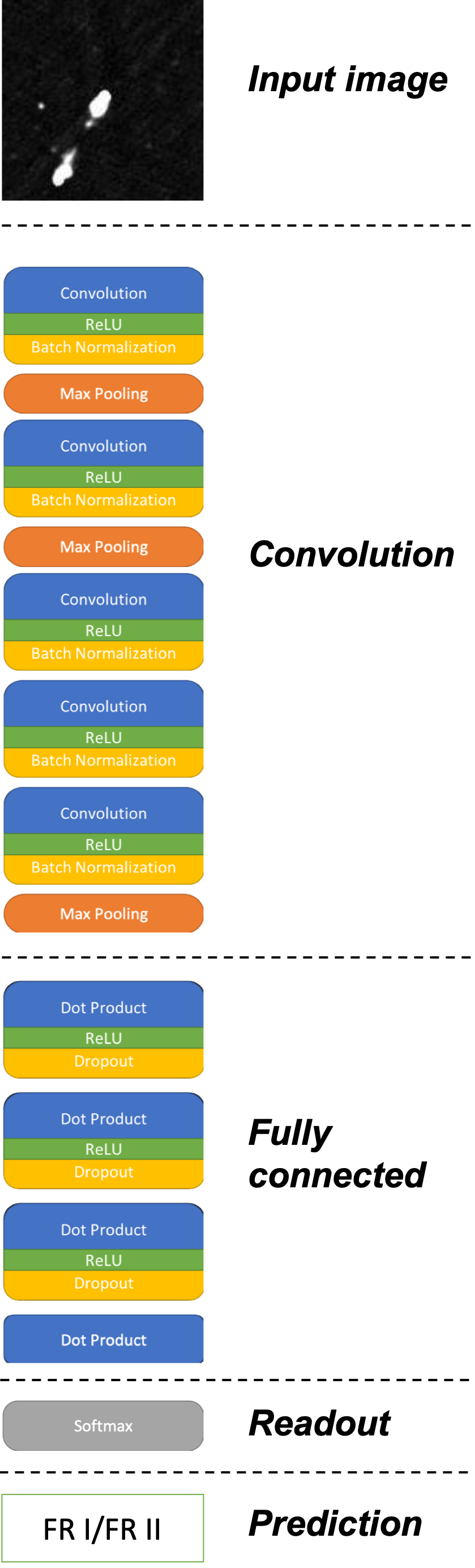}
\end{center}
\caption{Network architecture adopted in the work. Blue: filters with learnable parameters; Green: activation functions; Yellow: Regularizers; Orange: Pooling layers; Grey: Softmax layer. The 13-layer architecture contains 5 convolutional layers, 3 max-pooling layers, 4 fully-connected layers, and a softmax readout layer. We consider pooling and readout layers separately.}
\label{fig:Network_Graph}
\end{figure}

Inspired by \cite{2017ApJS..230...20A}, we adopt a similar but simplified 13-layer network 
see Fig.~\ref{fig:Network_Graph}. We initially discard the network components allowing for parallel computation. We then added another fully connected layer to reduce over-fitting. Finally, instead of optimizing loss using a traditional mini-batch gradient descent optimizer \citep[SGD;][]{Robbins1951} and step decay learning rate schedule, we used the adaptive mini-batch optimizer AdaGrad \citep{Duchi2011} to minimize the model loss function. This optimizer algorithm works with a batch size of 100 and a initial learning rate of 0.01. The batch size refers to the number of training samples fed into the network at a time. Since we use 22,268 training samples, we import our data in 223 batches. Such a data import method is often referred to as mini-batching. The mini-batching method has typically been found to speed up the training process \citep{Benetan2018}. 

Using this architecture, we observed that model validation accuracy started to saturate after 10-epoch training \citep{2017ApJS..230...20A}. Consequently, all training in this work is stopped after 10 epochs. The filter: node number for each layer is set to be 1:16 due to computational power limitations. The same architecture is used for both our initial model training and later transfer learning applications. Table~\ref{tab:CNN_Layers} provides further network details. Notably, the {\tt parameters} characteristic for each layer in Table~\ref{tab:CNN_Layers} is determined by the receptive field size, the input channel number, and the depth of each layer. The {\tt Conv1} layer, for instance, has $11 \times 11 \times 1 \times 6 + 6 = 732$ learnable parameters.

\begin{landscape}
\begin{table}
\begin{center}
\begin{tabular}{lllllllll}
\hline \hline
Layer   &  Name  & Receptive Field & Stride  & Input Channel Number & Depth &  Activation  & Regularizer & Parameters\\ \hline
1  & Conv 1  & 11 $\times$ 11  & 1   &   1   & 6  &  ReLU  & Batch Normalization  &  732    \\
2  & Max Pooling 1  & 2 $\times$ 2   & 2    & 6   &    &   &     &     \\
3  & Conv 2  & 5 $\times$ 5    & 1    &  6    &  16   &  ReLU  &  Batch Normalization  &  2,416   \\
4  & Max Pooling 2 & 3 $\times$ 3  & 3    &  16    &     &  &     &    \\
5  & Conv 3  & 3 $\times$ 3   & 1   &  16   & 24   &  ReLU  & Batch Normalization  &   3,456   \\
6  & Conv 4 & 3 $\times$ 3  & 1   &  24  &    24   & ReLU  &  Batch Normalization &   5,184      \\
7  & Conv 5   & 3 $\times$ 3    & 1      & 24      &   16   & ReLU   &  Batch Normalization &  3,456  \\
8  & Max Pooling 3  & 5 $\times$ 5   & 5   &   &    &   &    &   \\
9  & Fully Connected 1    &      &  1   & 5 $\times$ 5 $\times$ 16 & 256        & ReLU  & Dropout   &  102,656   \\
10 & Fully Connected 2    &      &    & 256 &  256   & ReLU  & Dropout  &   65,792   \\
11 & Fully Connected 3    &      &    & 256 &  256   & ReLU  & Dropout  &   65,792   \\
12 & Fully Connected 4    &      &    & 256 &  2   &   &   &   514   \\
13 & Loss Softmax    &                   &          &            &           &  &     &            \\ \hline 
Total Parameters: &   &                   &          &            &           &  &     &  249,998  \\ \hline \hline
\end{tabular}%
\caption{\label{tab:CNN_Layers} Network parameters of the classifier we adopted in the work. $\rm 'Parameters'$ are only available for $'\rm Conv'$ and FC layers. Parameters within these layer can learn through back propagation, while pooling and loss layers cannot learn.} 
\end{center}
\label{network_parameters_summary}
\end{table}
\end{landscape}


\subsection{Direct Classification}
\label{sec3_1}






\begin{figure}
\setlength{\unitlength}{1cm}
\begin{center}
\includegraphics[width=0.5\textwidth]{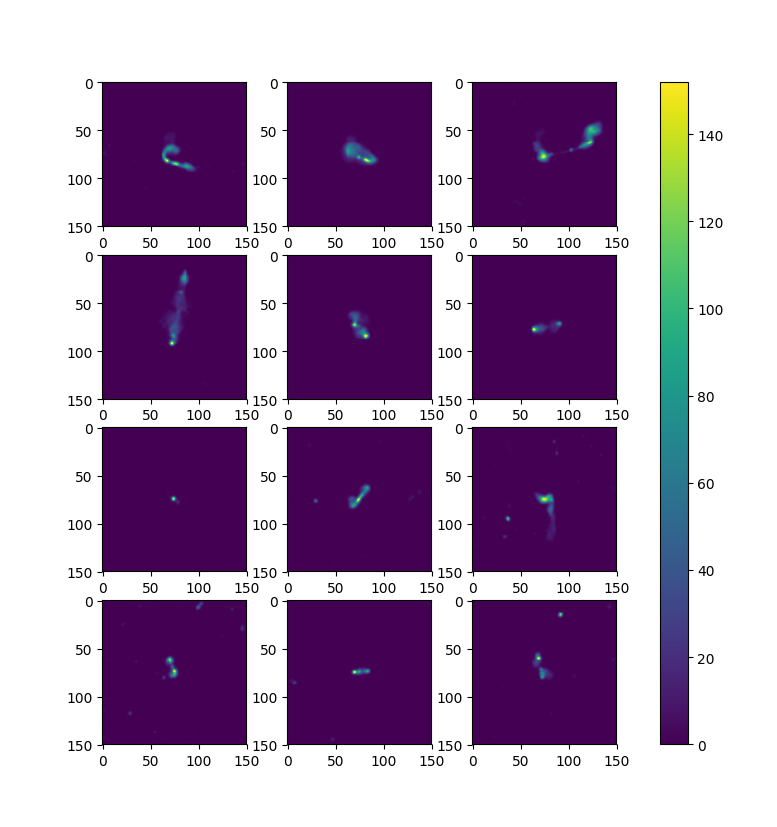}
\end{center}
\caption{Examples of images used in model training. Models trained with these samples were used to classify FR morphology from test dataset NVSS or FIRST images. 1st row: FR~I samples of FIRST images; 2nd row: FR~II samples of FIRST images; 3rd row: FR~I samples of NVSS images; 4th row: FR~II samples of NVSS images. The color bar represents the linear-normalized pixel values.}
\label{fig:png_example}
\end{figure}

CNNs are capable of learning the form of the features for classification and expressing them as the values of the convolution kernel weight matrix. In the context of radio morphology classification, CNNs will learn these features from the input training samples and extract differentiable features from them. Fig~\ref{fig:png_example} shows some typical examples of training samples. 

\begin{figure*}
\setlength{\unitlength}{1cm}
\begin{center}
\includegraphics[width=0.8\textwidth]{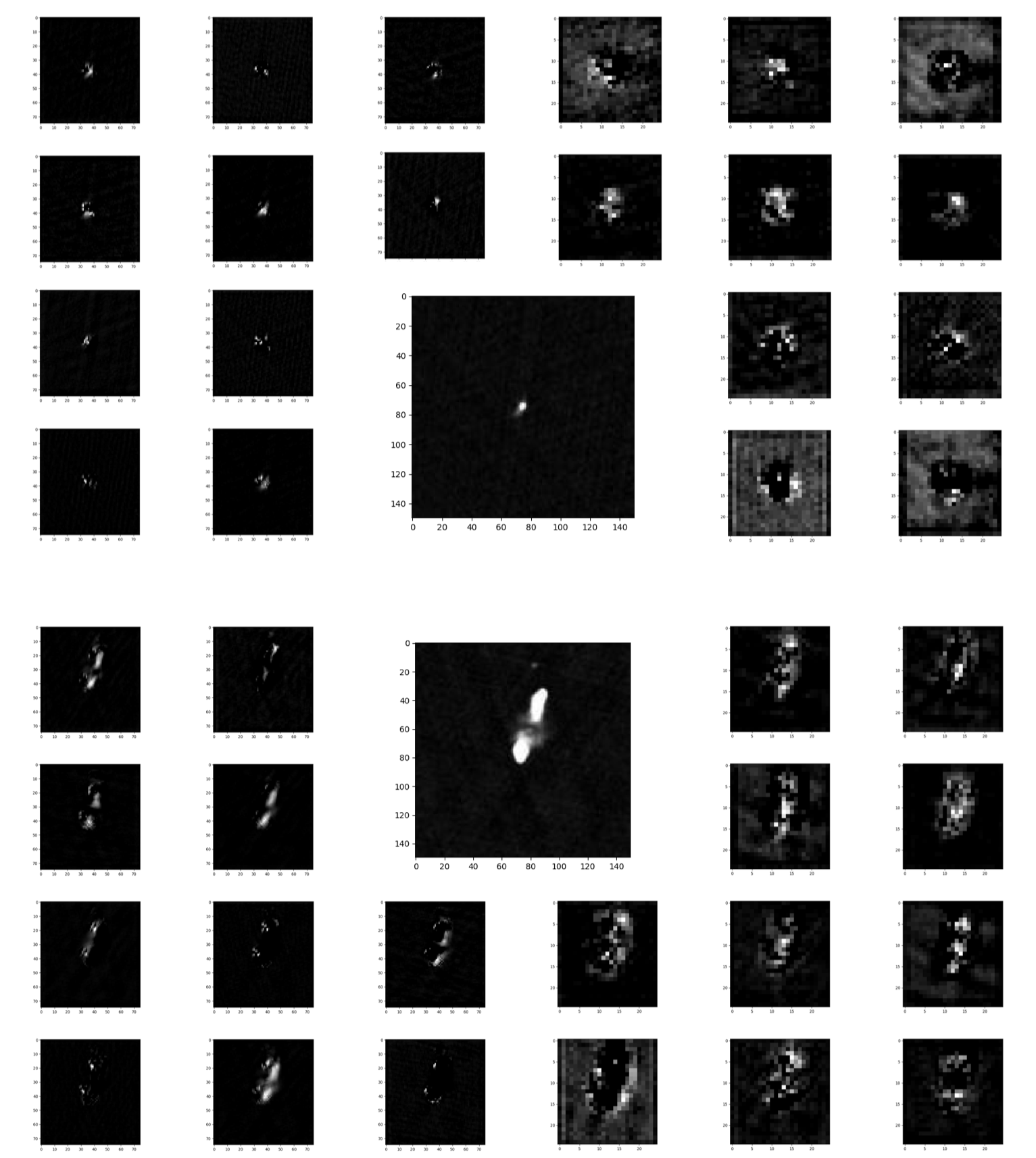}
\end{center}
\caption{ An example of feature maps using testing FIRST sample images. These images are produced by convolving the example source image with the first 10 filters of either the second or the fifth convolutional layer shown in Fig. \ref{fig:Network_Graph}. Upper-middle: An example of FR~I sources in the testing set. Lower-middle: An example of FR~II sources in the testing set. Upper-left: Features of the example FR~I source extracted by the second convolutional layer. Upper-right: Features of the example FR~I source extracted by the fifth convolutional layer. Lower-left: Features of the example FR~II source extracted by the second convolutional layer. Lower-right: Features of the example FR~II source extracted by the fifth convolutional layer. Source images and feature maps in the diagram are in grayscale. 
}

\label{fig:Feature_map}
\end{figure*}

When a convolutional network trains on these samples, its initial convolutional layers will tend to learn general sample features, while lower layers are more likely to extract features specific to the dataset itself. We visualize these features by plotting feature maps, which show the activation of different parts of the image \citep{2013arXiv1311.2901Z}. Fig~\ref{fig:Feature_map} shows feature maps for two representative samples. Features in the diagram were extracted from a randomly initialized model following 10-epoch training. These features correspond to the 2nd and the 5th convolutional layers. Both layers implement 16 filters in total and the figure shows the first 10. Generally speaking, for the FR~II source (bottom), the features learned by the 2nd layer seemed to emphasize the existence of double edge-brightened lobes and the relative positions of the hotspots. Whereas the lower layers have learned more specific features: source outlines or the source-background relationship. This is similar to what was observed in  \cite{2017ApJS..230...20A}. All these features are saved via their model weights, and are used by the fully connected layers to make an FR binary classification. 


\begin{figure*}
\setlength{\unitlength}{1cm}
\begin{center}
\includegraphics[width=\textwidth]{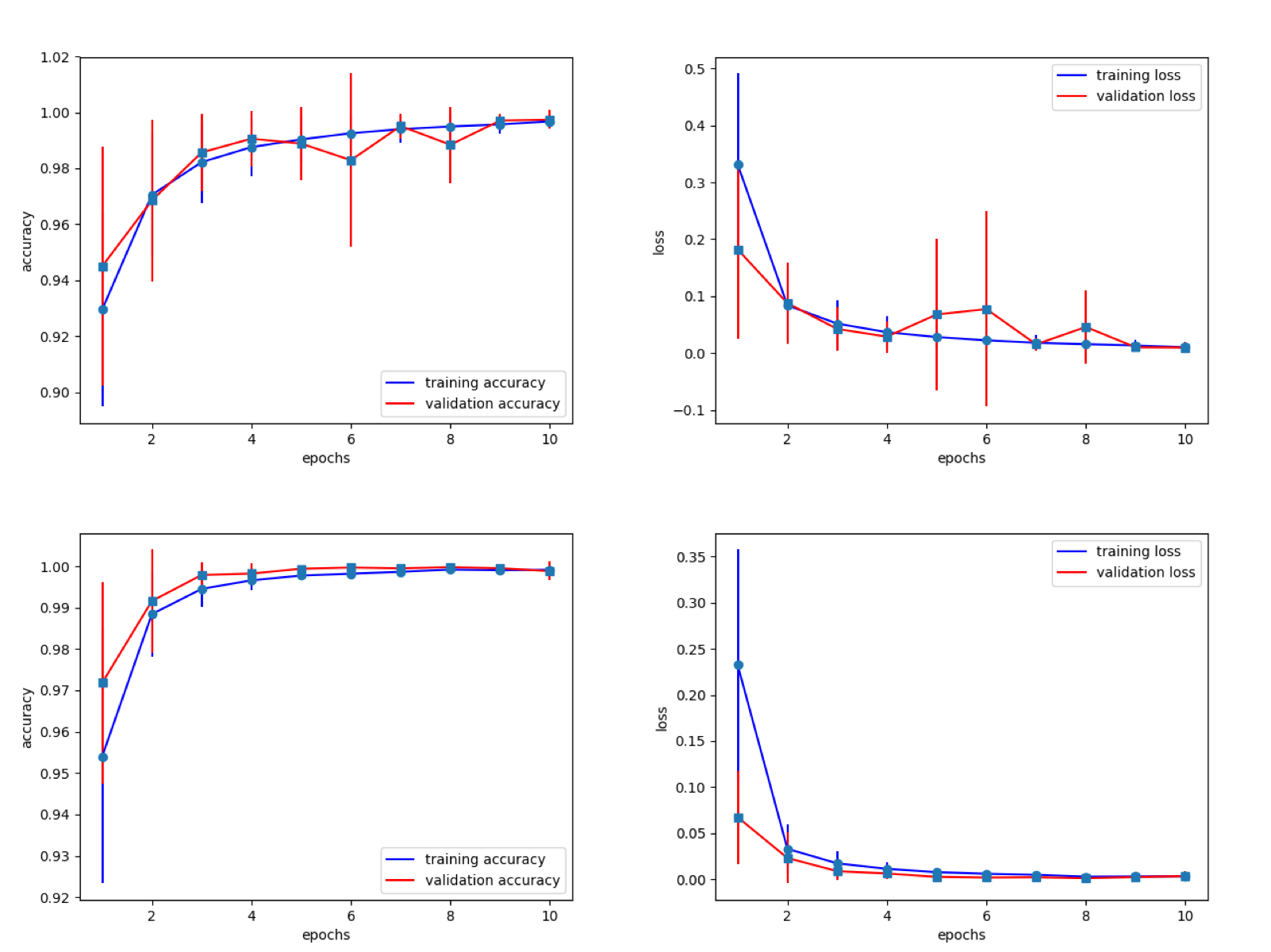}
\end{center}
\caption{Upper: averaging learning and loss curve for `Xavier' models trained on NVSS images. Lower: The same curves for models trained on FIRST images.}
\label{fig:Network_train_val_learning_loss_Scratch}
\end{figure*}

Both NVSS and FIRST training samples were imported to train a randomly initialized model for 10 epochs each. Fig~\ref{fig:Network_train_val_learning_loss_Scratch} 
provides an overview of model learning processes. The average temporary accuracy and loss is calculated from 10 independently trained models. The standard deviation of these parameters corresponds to their error bars in the diagram. All models trained from random initializations using the Xavier uniform weight initializer \citep[`Xavier' models hereinafter;][]{Xavier2010} that experienced 10 epochs of training have a validation accuracy above 99.5$\%$ and losses lower than 0.02, regardless of input sample selection.

Models trained using FIRST data as the input samples tended to have a smaller error. These models also learned more efficiently, as their training losses dropped faster. Models trained using NVSS images as input samples seemed to oscillate more frequently, which implies that these models have a higher risk of making random guesses and a lower chance of performing stable training. These results may not only be due to the comparatively poor resolution and sensitivity of NVSS versus FIRST. Some other possible causes are discussed in Sec \ref{sec4_1}.

\subsection{Transfer Learning}
\label{sec3_2}
Though models trained from scratch perform reasonably well, whether valuable to apply transfer learning on these models had yet to be explored. There was saying that algorithms trained on data from one survey have to be trained from other surveys \citep{2017ApJS..230...20A}, while it is uncertain if this can be valid regardless of the choices which input survey as a starting point. Here we considered three methods:

\begin{itemize}
	\item \textbf{Method 0:} Inherit the complete network architecture and weights from the pre-trained models and re-train the full network. 
    
	\item \textbf{Method A:} Inherit the same network and weights for all layers and  re-train the fully connected layers.
    
	\item \textbf{Method B:} Inherit the same network and the weights for the convolutional layers, but re-train the fully connected layers from scratch.
\end{itemize}

Here {\it re-train} means inheriting models trained with data from one survey as the network initialization, and then optimize on data from another survey. For example, we trained a network from scratch with NVSS images and then re-trained the model with FIRST images afterward, and vice versa. The transfer learning methods adopted here do not change the network architecture, the early stopping criteria (10 epochs), or the training sample size and hyperparameters (e.g. batch size and primary learning rate). These methods only specify which layers to freeze, and whether to trigger the training using pre-trained weights. 

The purpose of Method 0 is to examine whether transfer learning can provide a better starting point and form a better-trained model. Methods A and B are intended to examine if directly inheriting features from pre-trained convolutional layers can produce comparably good results and reduce training time. 

\begin{figure}
\setlength{\unitlength}{1cm}
\begin{center}
\includegraphics[width=0.5\textwidth]{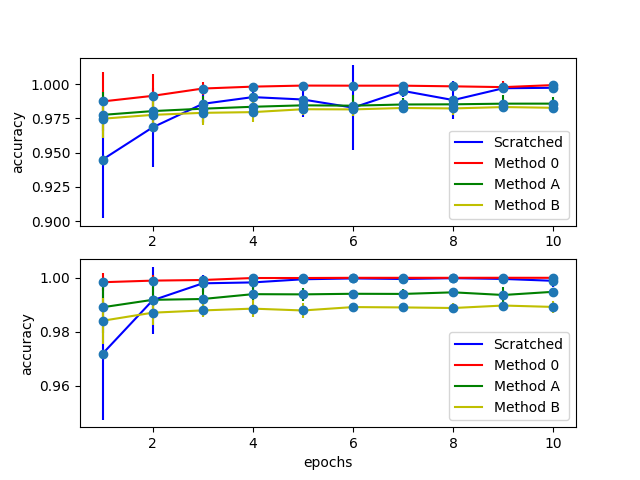}
\end{center}
\caption{Upper: Model average validation accuracy curves with corresponding error bars trained with inherit NVSS images using variant methods. Blue, red, green and yellow curves represent models trained from scratch, using Methods 0, A, and B, respectively. Lower: The same curves trained with inherited weights trained on FIRST images.}
\label{fig:Network_train_val_learning_loss}
\end{figure}

Fig~\ref{fig:Network_train_val_learning_loss} shows average learning curves for each of the three methods and includes the `Xavier' models for comparison. In general, transfer learning 
constrains the standard deviation compared to direct training by at least a factor of 2. Such an effect is seen regardless of the transfer learning method applied or survey data used. In addition, transfer learning accelerated the convergence of the model in each case. Finally, the application of transfer learning provided a higher starting validation accuracy after the first epoch of training. 

Among the three methods, Method 0 seems to perform best. By applying Method 0, models share highest starting points for training on both NVSS (98.7$\%$) and FIRST (99.8$\%$) after only one epoch of training. Though the accuracy gap between the `Xavier' models and the Method 0 models gradually decreases, the models using Method 0 in fact provide better validation accuracy with final values above 99.9$\%$. In the case of independent training stability, the `Xavier' models have a validation accuracy standard deviation after the final epoch on order of $10^{-3}$, whereas the value for models using Method 0 ranges from $10^{-5}$ to $10^{-4}$. The time spent on training for models using Method 0 is slightly longer than training from scratch: randomly initialized models need 33\,ms to train on an image, while the time required for Method 0 is about $\sim$34\,ms. Generally speaking, each training run takes $\sim$124\,minutes on our test system, see \S~\ref{sec0_0}. 

Comparatively, Methods A and B may require more training over a larger number of epochs as their training accuracies grow very slowly. Models using Method A or B have a validation accuracy which grows from 0.5$\%$ to 0.8$\%$ at each epoch, while the growth for `Xavier' models is $>2.8\%$. This is understandable given that the convolutional layers are frozen. However, freezing these layers reduces the time cost for model training. Both Methods A and B require only 7\,ms to train each input image, $\sim$21$\%$ of the time cost using Method 0. Comparing the two methods, Method A performs better that Method B due to the difference in its weight inheritance. Further analysis of classification accuracy using the three methods will be given in \S~\ref{sec4_1}.

\section{Discussion}
\label{sec4_0}




\subsection{Classification Accuracy}
\label{sec4_1}



Besides validation sets, the prediction-truth comparison is another factor when evaluating a classifier. Accuracy is often considered as the primary parameter for evaluation; however, whilst this metric can provide an overview of model performance, this can be misleading when test samples have a significant class imbalanced.

\begin{figure}
\setlength{\unitlength}{1cm}
\begin{center}
\includegraphics[width=0.5\textwidth]{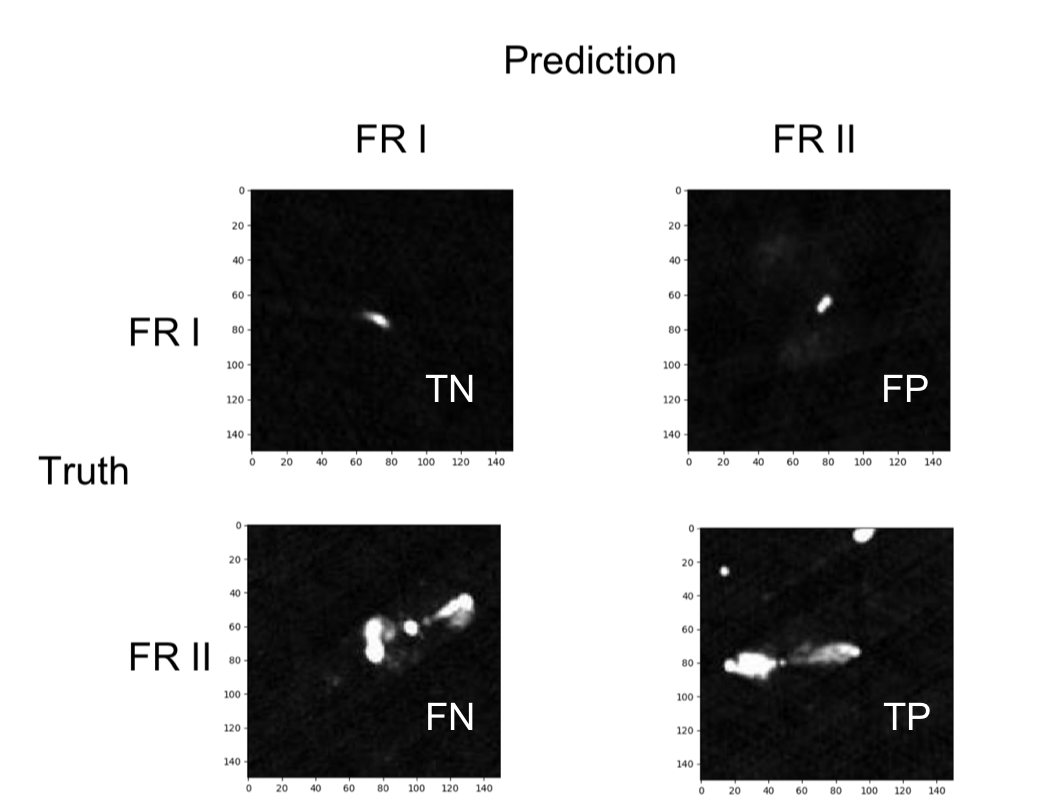}
\end{center}
\caption{An example of a confusion matrix. In the context of binary classification, FR~I and FR~II represent false and true classes. All pre-processed FIRST images in the matrix came from the test set.}
\label{fig:Confusion_matrix}
\end{figure}

A widely used tool for evaluating model performance on a class-wise basis is the confusion matrix \citep{Stehman1997}, a table to visualize model performance. Fig~\ref{fig:Confusion_matrix} gives an example of a confusion matrix adopted for this work. This matrix has four quadrants: True-Positive (TP), False-Negative (FN), True-Negative (TN), and False-Positive (FP). Assuming FR~II morphology as ``true," the matrix counts the FR~II test samples towards TP if model prediction matches its class label identified by human.

The four-quadrants can be used to further assess the predictive ability of a two class model as the basis for deriving three additional metrics: Recall ({\tt R}), Precision ({\tt P}), and $F_{1}$ score \citep{Powers2008}. {\tt R} is the ratio of TP and TP + FP, while {\tt P} is TP /(TP + FN). In the context of classifying FR~II morphology, a higher {\tt R} and {\tt P} score refers to greater sensitivity and class prediction accuracy, respectively. The $F_{1}$ score,
%
\begin{ceqn}
\begin{align}
\rm F_{1} = 2~ \frac{{\tt P} \times {\tt R}}{{\tt P} + {\tt R}},
\label{eqn:II}
\end{align}
\end{ceqn}
can be seen as the weighted average of {\tt R} and {\tt P}, which provides a general assessment when identifying samples of a class. These four metrics enable us to evaluate the models trained in this work. Each model was tested with identical NVSS and FIRST image sets. 

In addition to the numerical value for each of these four metrics, the Receiver Operating Characteristic (ROC; e.g., Fig~\ref{fig:Network_ROC}) curve is also used to  represent model performance. The ROC curve is mainly a useful tool when making binary classification, although multi-class variants do exist. It provides a visualisation of the  false-positive rate versus the  true-positive rate for a number of candidate thresholds from 0 to 1. The true-positive rate is equal to the recall when the class being considered is the `real' class, while Equation~\ref{eqn:III} defines the false-positive rate:
%
\begin{ceqn}
\begin{align}
\rm False~positive~rate = \frac{{\tt fp}}{{\tt fp} + {\tt tn}}.
\label{eqn:III}
\end{align}
\end{ceqn}

The ROC curve can be seen as a trade-off between the two variables, and the area under the curve (AUC) value is often used when comparing models. In such a comparison, testing on the same samples, the model with a higher AUC is considered to perform best when distinguishing classes.

\subsubsection{Randomly initialized models}
\label{sec4_1_1}
\begin{figure}
\setlength{\unitlength}{1cm}
\begin{center}
\includegraphics[width=0.5\textwidth]{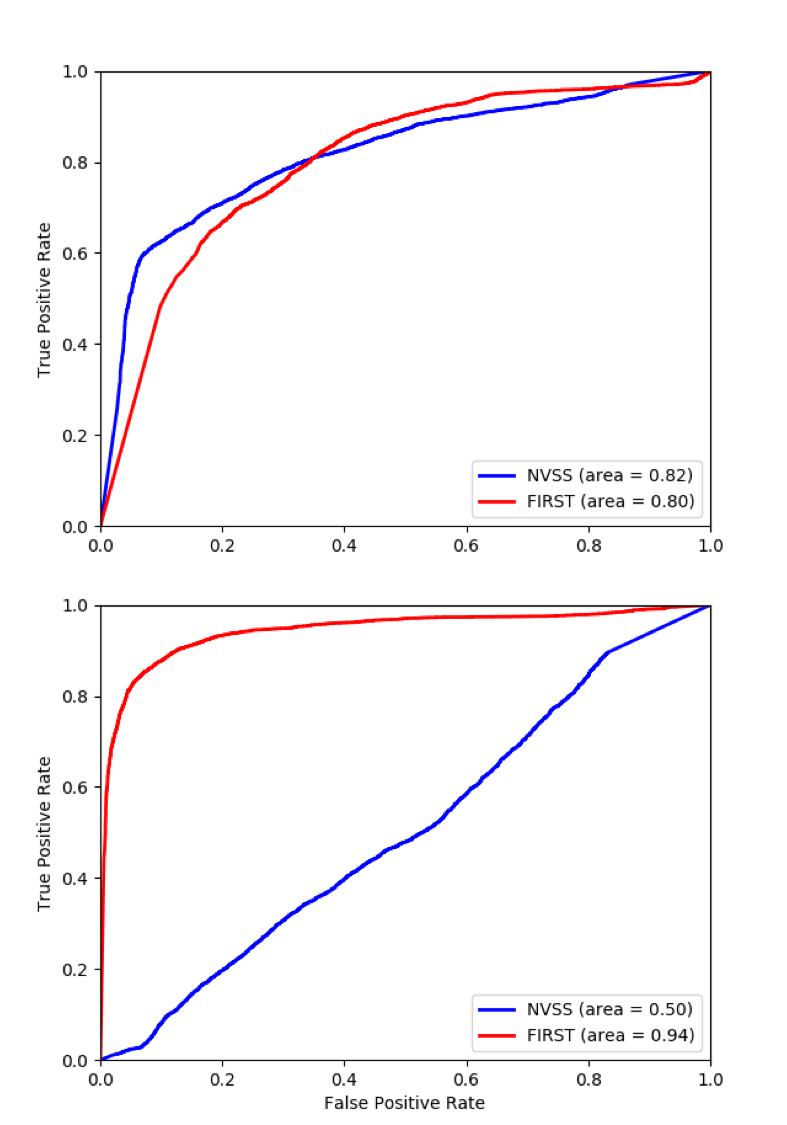}
\end{center}
\caption{ROC curve for `Xavier' models. The colors in the diagram represent the survey of the test images used to derive the curve. Blue refers to NVSS images, while red represents FIRST images. Upper: ROC curves for `Xavier' models trained on NVSS images for 10 epochs. Lower: ROC curves for `Xavier' models trained on FIRST images. When deriving the curves, the FR~I class is assumed to be ``true", while the FR~II class is considered to be ``false''. 
} 
\label{fig:Network_ROC}
\end{figure}

Depending on the choice of datasets used in training and testing, classifier performance can vary under evaluation. Fig~\ref{fig:Network_ROC} shows the ROC curves for different models trained from scratch. Models trained with NVSS samples show similar behavior when tested on either NVSS or FIRST samples. The average AUC for testing the two sample sets is 0.80 $\pm$ 0.01 and 0.78 $\pm$ 0.02, respectively.

In comparison, models trained using FIRST samples show an asymmetric performance. Such models work well when classifying unlabelled (test) FIRST images, but they behave randomly when tested on NVSS images. The AUC for these models reached 0.94 $\pm$ 0.01 for the FIRST samples, while the metric for NVSS was 0.54 $\pm$ 0.05. 

For other metrics the situation is similar. Table~\ref{tab:Scratched model performance} summarizes the metrics described above for these models. It can be seen that models with higher AUC scores also share higher classification accuracy. Models trained using FIRST images perform best when making predictions on FIRST test images. These models generally achieve $89.1\% \pm 1.4\%$ accuracy. When models are trained and tested on NVSS images, however, test accuracy drops to $73.0\% \pm 1.1\%$. Such a change might be attributed to the differences between the two surveys: sample sources in FIRST are well resolved and extended in most cases, sources in NVSS, however, are sometimes only slightly resolved and are small in the image (Fig~\ref{fig:png_example}).

When models were trained on FIRST images and tested on NVSS images, however, we saw strong FR~I class preference in these models. Neither model recall nor precision when classifying NVSS FR~II images is higher than 0.5. Such bias also exists when models are trained with NVSS images. All randomly initialized models perform better when classifying NVSS FR~I samples. Given that the training data is well balanced, with a 0.45$\%$ higher number of FR~I samples than FR~IIs, it is unlikely that this is a consequence of class imbalance alone.


Since the model trained by \cite{2017ApJS..230...20A} used similar input data samples, we naively compare the results of our randomly initialized models trained on FIRST images to that work. Our precision when classifying FR~I objects is $\sim95\%$, similar to theirs. For FR~II classification, our models achieved 83$\%$ precision, compared to 75$\%$ from their models. The average F1 score in our work is 91$\%$, 5$\%$ higher than \cite{2017ApJS..230...20A}; however, the recall of their models for classifying FR~Is on the other hand is 6$\%$ higher than ours.  The difference between these two works may be explained in several ways. 

Firstly, the fusion model they proposed was a three-class classification model and a number of bent-tailed radio galaxies were mistakenly identified as FR~II sources \citep{2017ApJS..230...20A}. Secondly, there is a discrepancy in the number of input data samples and their distribution. When training their models, \citep{2017ApJS..230...20A} imported 36,000 FR~Is, 32,688 FR~IIs, and 25,488 Bent-Tailed `sources', whereas our complete data sample contains 39,796 samples. This might contribute to higher recall. Finally, although we imported FR~Is and FR~IIs from the same catalogs, the definition of FR~Is between the two differs slightly: some FR~Is we imported were considered as bent-tailed sources \citep{2017ApJS..230...20A}. This could potentially cause a difference in model performance.

\begin{table}
\begin{center}
\begin{tabular}{| c | c | c | c | c |}
\hline \hline
\multirow{1}{*}{\textbf{NVSS}} & \multicolumn{2}{ c |}{\textbf{NVSS test result}} & \multicolumn{2}{ c |}{\textbf{FIRST test result}}   \\ 

\textbf{trained} & \multicolumn{1}{ c |}{\textbf{FR~I}}   & \textbf{FR~II}  & \multicolumn{1}{ c |}{\textbf{FR~I}} & \textbf{FR~II} \\
\hline
Recall & \multicolumn{1}{ c |}{0.67$\pm$0.01}   & 0.87$\pm$0.04  & 0.74$\pm$0.06  & 0.70$\pm$0.06  \\ 
Precision & \multicolumn{1}{ c |}{0.92$\pm$0.02}   & 0.54$\pm$0.03  & 0.67$\pm$0.04 & 0.77$\pm$0.07 \\ 
F1 Score & \multicolumn{1}{ c |}{0.77$\pm$0.02}   & 0.67$\pm$0.05 & 0.70$\pm$0.08 &  0.73$\pm$0.10 \\
Accuracy($\%$) & \multicolumn{2}{ c |}{73.0$\pm$1.1}   & \multicolumn{2}{ c |}{71.9$\pm$2.8}    \\
\hline
\multirow{ 1}{*}{\textbf{FIRST}} & \multicolumn{2}{ c |}{\textbf{NVSS test result}} & \multicolumn{2}{ c |}{\textbf{FIRST test result}}  \\ 

\textbf{trained}& \multicolumn{1}{ c |}{\textbf{FR~I}} & \textbf{FR~II} & \multicolumn{1}{ c |}{\textbf{FR~I}} & \textbf{FR~II}   \\
\hline
Recall & \multicolumn{1}{ c |}{0.49$\pm$0.01} & 0.40$\pm$0.17  & 0.85$\pm$0.02 & 0.94$\pm$0.04 \\ 
Precision & \multicolumn{1}{ c |}{0.92$\pm$0.02}  & 0.05$\pm$0.02  & 0.95$\pm$0.02 & 0.83$\pm$0.04 \\ 
F1 Score & \multicolumn{1}{ c |}{0.64$\pm$0.02}  & 0.09$\pm$0.06  & 0.90$\pm$0.03 & 0.88$\pm$0.06 \\ 
Accuracy($\%$) & \multicolumn{2}{ c |}{48.5$\pm$1.2}   & \multicolumn{2}{ c |}{89.1$\pm$1.4}    \\
\hline \hline
\end{tabular}
\end{center}
\caption{A summary of model performance for randomly initialized models trained for 10 epochs. Testing for models trained on one survey images adopted the test image set from the same survey.}
\label{tab:Scratched model performance}
\end{table}

\subsubsection{Transfer learning models}
\label{sec4_1_2}
Although a naive analysis of the transfer learning models presented in this work might initially indicate the advantages of applying these methods, it is important to consider that these models may show varying performance characteristics when applied to new or different unseen datasets. 
Table~\ref{tab:transfer_leraning_accuracy} gives a summary of the test accuracy for models trained with or without transfer learning methods, when applied to a dataset different from that used for original training. Model test accuracy represents the general performance of a classifier.

The application of Method 0 boosted model classification ability when predicting NVSS images and gave the best test performance among the three methods. When classifying FIRST images, however, models which inherited pre-trained weights from FIRST images and applied Method A performed best. The same result is not true in the case where the order of the survey data used for the inherited-retraining sequence was switched from FIRST to NVSS.  

The AUC values also provide further detail. Models which used Method 0 showed comparatively stronger expressive ability than randomly initialized ones. Adopting Methods A and B produced a similar effect when inheriting weights from models pre-trained on FIRST images. Such a phenomenon, however, lost its efficacy if inheriting weights from models initially trained on NVSS images. This can perhaps be explained by the difference between the two surveys. Images of many sources seen in the FIRST survey possess richer structural information than provided by NVSS. This is an important factor when considering the application of pre-trained models from existing surveys to new data from next-generation telescopes such as ASKAP, MeerKAT and the SKA.

\begin{table}
\begin{center}
\begin{tabular}{| c | c | c | c | c |}
\hline \hline
\multirow{1}{*}{\textbf{NVSS}} &   &    \\ 

\textbf{trained} & \multicolumn{1}{ c |}{\textbf{Xavier}}   & \textbf{0}  & \multicolumn{1}{ c |}{\textbf{A}} & \textbf{B} \\
\hline
NVSS test($\%$) & 73.0$\pm$1.1   & \textbf{73.0$\pm$0.7} & 69.8$\pm$1.0  & 71.6$\pm$0.8   \\
FIRST test($\%$) & 71.9$\pm$2.8   & 78.4$\pm$2.0 & \textbf{81.1$\pm$1.0} & 78.3$\pm$1.0   \\
\hline
\multirow{ 1}{*}{\textbf{FIRST}} &  &   \\ 

\textbf{trained}& \multicolumn{1}{ c |}{\textbf{Xavier}} & \textbf{0} & \multicolumn{1}{ c |}{\textbf{A}} & \textbf{B}   \\
\hline
NVSS test($\%$) & 48.5$\pm$1.2   & \textbf{50.3$\pm$0.7} & 46.9$\pm$0.5 & 46.2$\pm$0.6   \\
FIRST test($\%$) & \textbf{89.1$\pm$1.4}  & 87.4$\pm$1.4  & 84.6$\pm$0.6 & 83.8$\pm$1.0   \\
\hline \hline
\end{tabular}
\end{center}
\caption{A summary of averaging model accuracy. Accuracy in the table are represented in percentage. 'trained' refers to the survey data finally trained on each model. Bold implies that the method horizontally gave the best accuracy.}
\label{tab:transfer_leraning_accuracy}
\end{table}

\begin{figure}
\setlength{\unitlength}{1cm}
\begin{center}
\includegraphics[width=9cm]{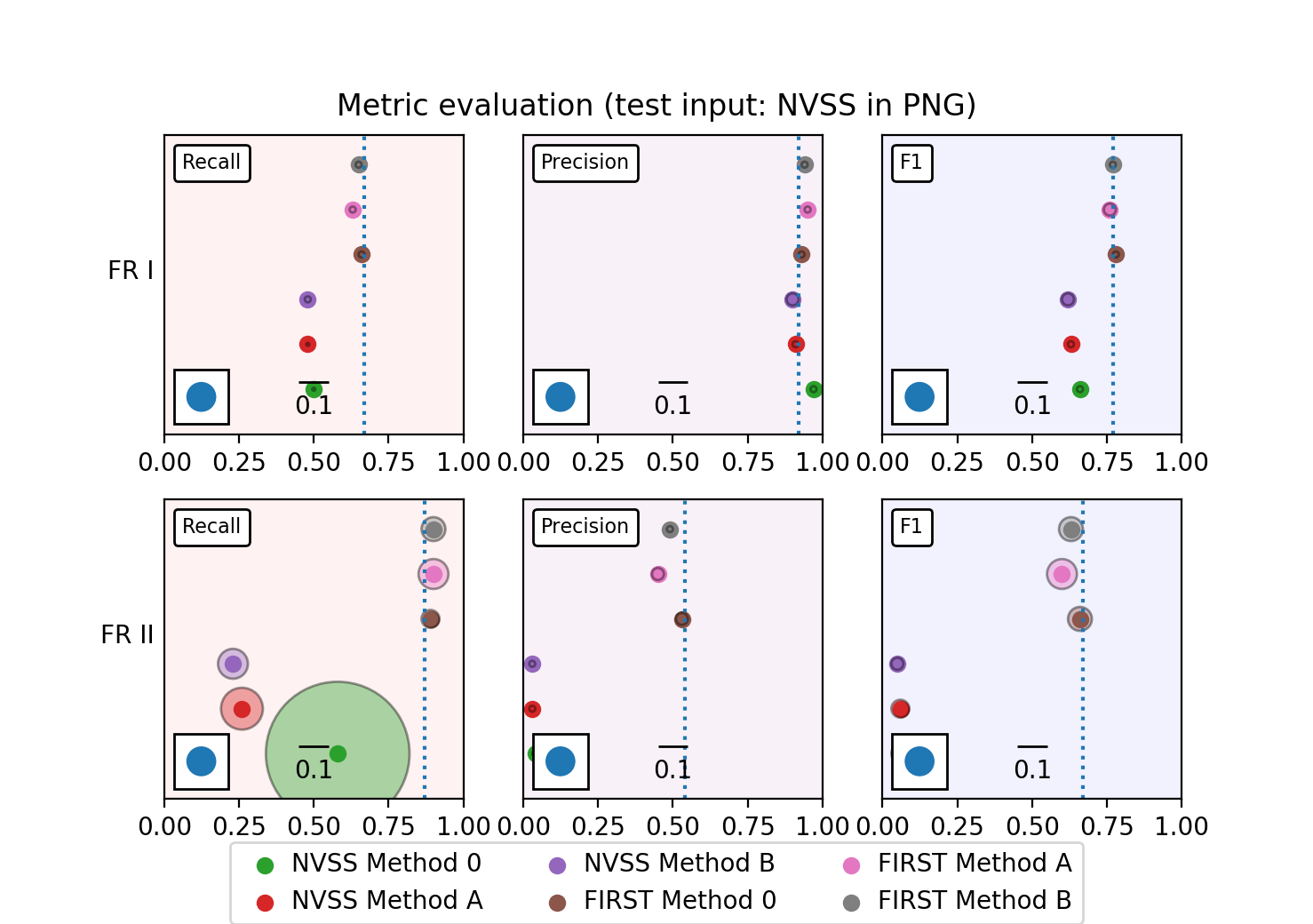}
\end{center}
\caption{A summary of metric evaluation for models applied transfer learning and tested on NVSS images. `NVSS' or `FIRST' shown in the legend box implies that, when applying transfer learning, the pre-trained model weights were trained on the named survey. In the diagram, radius of the circles accounts for the standard deviations of their respective metrics. Dashed vertical lines refer to average metrics for the Xavier models trained and tested on NVSS images.}
\label{fig:Metric_test_NVSS}
\end{figure}

\begin{figure}
\setlength{\unitlength}{1cm}
\begin{center}
\includegraphics[width=9cm]{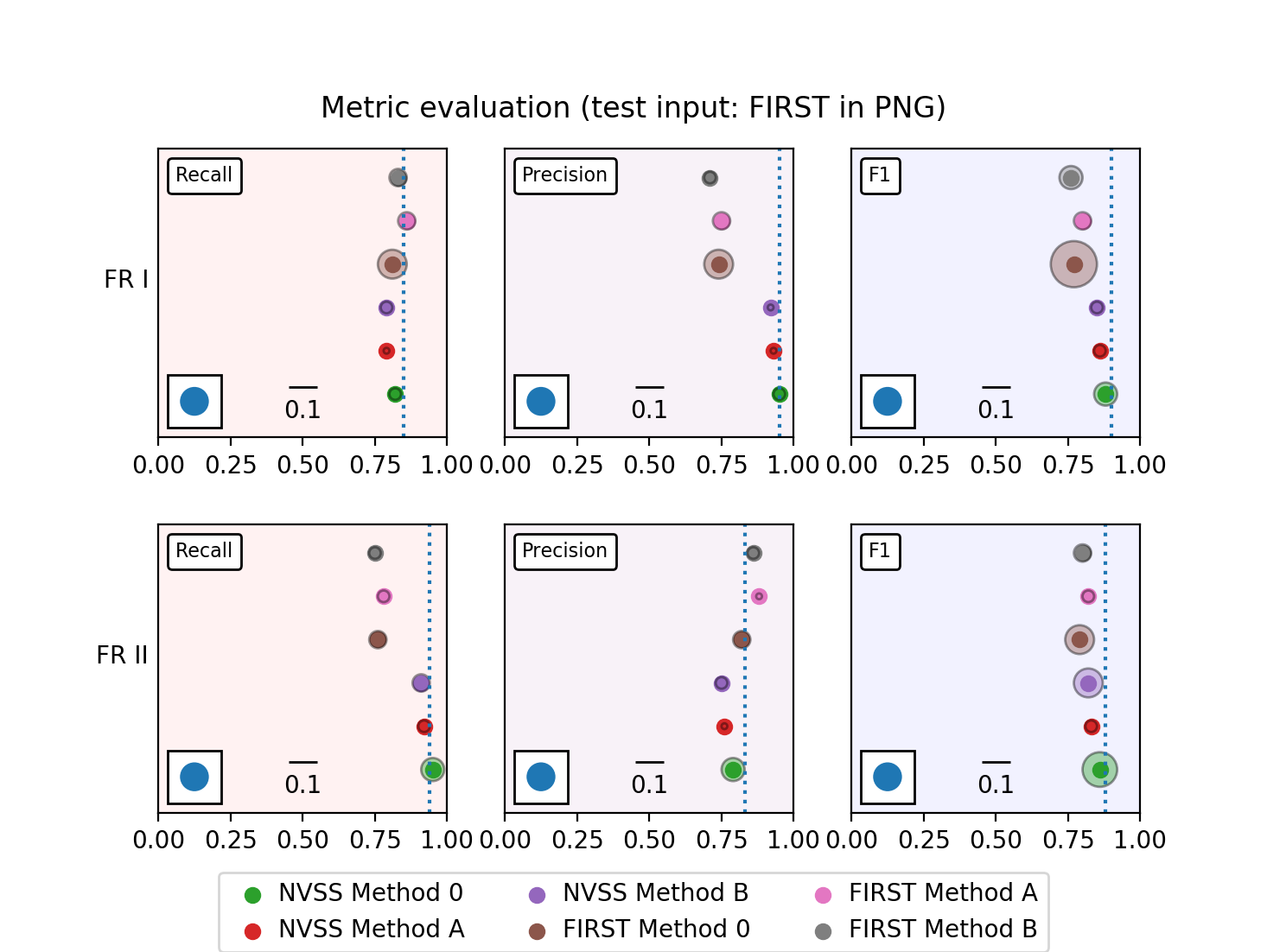}
\end{center}
\caption{A summary of metric evaluation for models applied transfer learning and tested on FIRST images. Models evaluated in the diagram are the same as Figure~\ref{fig:Metric_test_NVSS}. The meanings of symbols and texts in the diagram are consistent to Figure~\ref{fig:Metric_test_NVSS} as well. Dashed vertical lines refer to average metrics for the Xavier models trained and tested on FIRST images.}
\label{fig:Metric_test_FIRST}
\end{figure}

In addition to accuracy and AUC values, we also evaluated these transfer learning models further by measuring their recall, precision, and F1 score for each class. Figs~\ref{fig:Metric_test_NVSS}~\&~\ref{fig:Metric_test_FIRST} show how transfer learning models behave when classifying either FR~Is or FR~IIs on NVSS and FIRST images. We note that models consistently identified most test samples as FR~Is if they were re-trained with FIRST inputs and tested using an NVSS test dataset. When models are re-trained with NVSS images we found that introducing transfer learning improved both FR~I and FR~II classification for NVSS images. The identification of FR~II objects typically reached 88$\%$ precision using Method A. This is $\sim 5\%$ higher than models trained with FIRST images directly. In the context of FR~I classification using FIRST samples, the highest achieved precision value was 95$\%$. Such precision could either be achieved through direct training or by using Method 0 when training with FIRST images. 

These results imply that choice of method is a trade-off. Applying Method 0 would strengthen model stability and boost the performance of a model if one wished to apply the model to both NVSS and FIRST images. Method A can boost prediction precision when identifying FR~IIs on FIRST images. If not training FIRST images from scratch, applying Method A can make the most precise prediction on FIRST images. Method B can be seen as an alternative option in the case where computational power is constrained and one wants a quickly trained model which makes a reasonably good prediction. 

\begin{table}
\begin{center}
\begin{tabular}{| c | c | c | c | c |}
\hline \hline
\multirow{1}{*}{\textbf{NVSS}} &   &    \\ 

\textbf{trained} & \multicolumn{1}{ c |}{\textbf{Xavier}}   & \textbf{0}  & \multicolumn{1}{ c |}{\textbf{A}} & \textbf{B} \\
\hline
NVSS test & 0.80$\pm$0.01   & \textbf{0.81$\pm$0.01}   & 0.80$\pm$0.01  & 0.81$\pm$0.00  \\
FIRST test & 0.78$\pm$0.02   & 0.86$\pm$0.01 & \textbf{0.88$\pm$0.01} & 0.83$\pm$0.01   \\
\hline
\multirow{ 1}{*}{\textbf{FIRST}} &  &   \\ 

\textbf{trained}& \multicolumn{1}{ c |}{\textbf{Xavier}} & \textbf{0} & \multicolumn{1}{ c |}{\textbf{A}} & \textbf{B}   \\
\hline
NVSS test & 0.54$\pm$0.05   & \textbf{0.59$\pm$0.02}  & 0.57$\pm$0.01 & 0.53$\pm$0.01   \\
FIRST test & 0.94$\pm$0.01  & \textbf{0.94$\pm$0.00}  & 0.93$\pm$0.00 & 0.92$\pm$0.00   \\
\hline \hline
\end{tabular}
\end{center}
\caption{A summary of averaging model AUC. `trained' refers to the survey data finally trained on each model. Bold implies that the method horizontally gave the highest AUC.}
\label{tab:transfer_leraning_auc}
\end{table}

\subsubsection{Influence of input image format}
\label{sec4_1_3}
The input images used for model training and testing described in \S~\ref{sec4_1_1} and \S~\ref{sec4_1_2} are processed in PNG image format. When saving images, such a format converts the value of each pixel to an integer in a lossless fashion. 

In addition to PNG format, many classifiers also accept images in JPEG format. The advantage of using JPEG images is that they can require smaller storage volume and have enhanced smoothness. However, when converting image arrays to JPEG format, the images are compressed and there is information loss.  

The influence of input image format on model performance has not been addressed in the context of radio galaxy classification. However, the issue of archival data storage for the next generation of radio telescopes may have implications for training data availability. In order to investigate the effect of image format we repeated the image pre-processing, data augmentation, model training, and transfer learning processes described above using images input in JPEG format in order to compare the resulting model outcomes with those using PNG inputs.

\begin{table}
\begin{center}
\begin{tabular}{| c | c | c | c | c |}
\hline \hline
\multirow{1}{*}{\textbf{NVSS}} &   &    \\ 

\textbf{trained} & \multicolumn{1}{ c |}{\textbf{Xavier}}   & \textbf{0}  & \multicolumn{1}{ c |}{\textbf{A}} & \textbf{B} \\
\hline
NVSS test($\%$) & 82.1$\pm$4.4   & \textbf{83.9$\pm$1.2} & 81.8$\pm$0.6  & 82.6$\pm$0.6   \\
FIRST test($\%$) & 65.5$\pm$7.4   & \textbf{73.7$\pm$2.0} & 65.9$\pm$1.0 & 66.0$\pm$1.0   \\
\hline
\multirow{ 1}{*}{\textbf{FIRST}} &  &   \\ 

\textbf{trained}& \multicolumn{1}{ c |}{\textbf{Xavier}} & \textbf{0} & \multicolumn{1}{ c |}{\textbf{A}} & \textbf{B}   \\
\hline
NVSS test($\%$) & \textbf{57.8$\pm$1.9}   & 57.7$\pm$1.2 & 56.6$\pm$0.5 & 57.2$\pm$1.7   \\
FIRST test($\%$) & \textbf{90.1$\pm$1.0}  & 89.7$\pm$0.8  & 87.6$\pm$0.7 & 87.2$\pm$0.9   \\
\hline \hline
\end{tabular}
\end{center}
\caption{A summary of averaging model accuracy. Model inputs considered in the diagram are in JPEG image format. Accuracy in the table are represented in percentage. 'trained' refers to the survey data finally trained on each model. Bold implies that the method horizontally gave the best accuracy.}
\label{tab:transfer_leraning_accuracy_JPEG}
\end{table}

Table~\ref{tab:transfer_leraning_accuracy_JPEG} summarizes model performances using image inputs in JPEG format. Models using JPEG inputs show stronger identification ability. Comparing with Table~\ref{tab:transfer_leraning_accuracy}, classifiers primarily trained with NVSS images showed a 9$\%$ accuracy improvement when classifying NVSS test sets. For those models trained with FIRST images, however, classification accuracy is boosted for both NVSS and FIRST test datasets. 

\begin{figure}
\setlength{\unitlength}{1cm}
\begin{center}
\includegraphics[width=9cm]{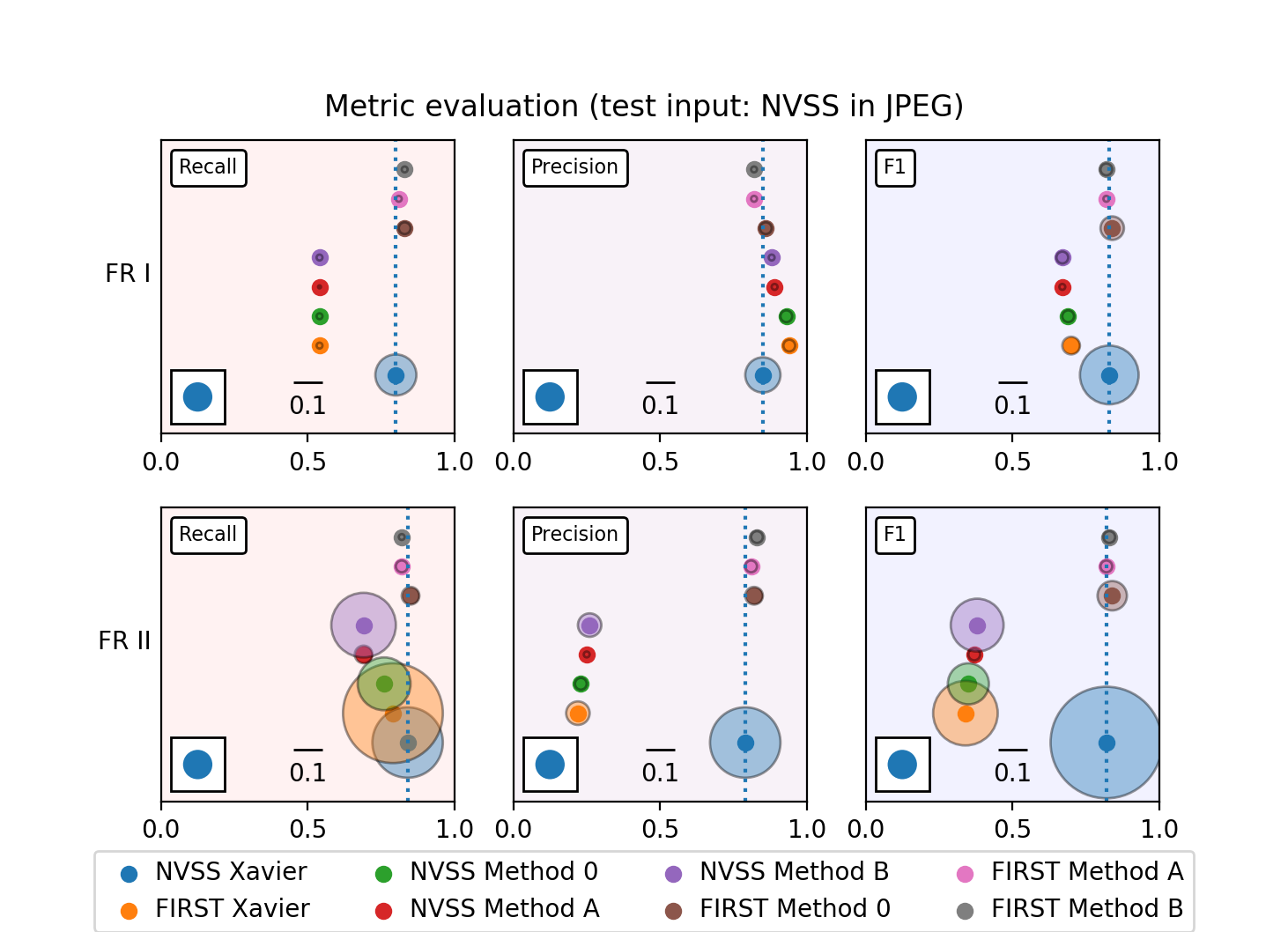}
\end{center}
\caption{A summary of metric evaluation for models applied transfer learning and tested on NVSS images in JPEG input format. For transfer learning models, 'NVSS' or 'FIRST' shown in the legend box implies that, the pre-trained model weights were trained on the named survey. For `Xavier' models, however, survey name refer to the survey data used in model training. In the diagram, the radius of the circles accounts for the standard deviations of their respective metrics. Dashed vertical lines, on the other hand, represent the average metrics for Xavier models trained and tested on NVSS images.}
\label{fig:Metric_test_NVSS_JPEG}
\end{figure}

\begin{figure}
\setlength{\unitlength}{1cm}
\begin{center}
\includegraphics[width=9cm]{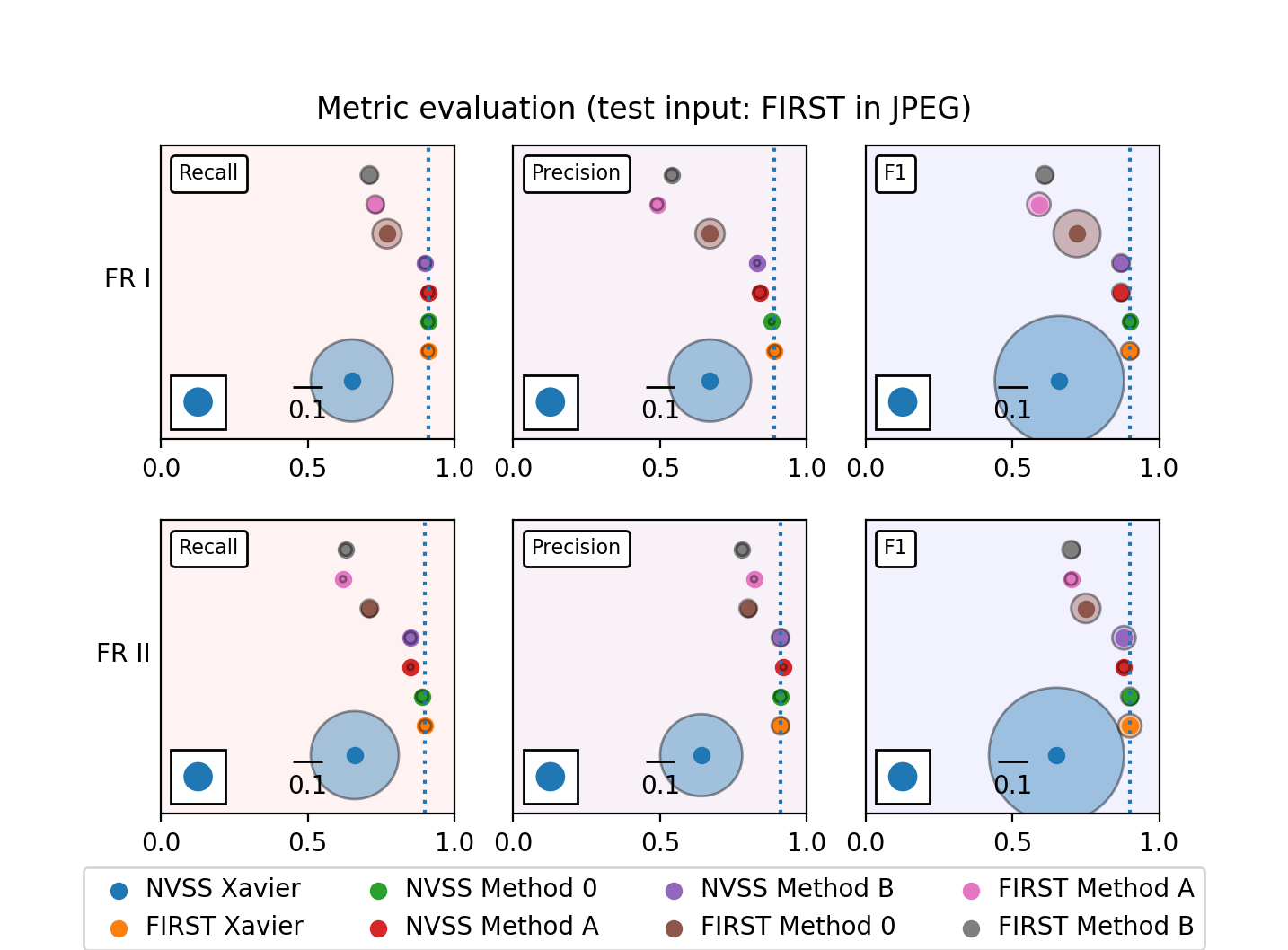}
\end{center}
\caption{A summary of metric evaluation for models applied transfer learning and tested on FIRST images in JPEG input format. Models evaluated in the diagram are the same as Figure~\ref{fig:Metric_test_NVSS_JPEG}. The meanings of symbols and texts in the diagram are consistent to Figure~\ref{fig:Metric_test_NVSS_JPEG}. Dashed vertical lines, on the other hand, represent the average metrics for `Xavier' models trained and tested on FIRST images.}
\label{fig:Metric_test_FIRST_JPEG}
\end{figure}

When considering Figures~\ref{fig:Metric_test_NVSS_JPEG}~\&~\ref{fig:Metric_test_FIRST_JPEG}, it can be seen that the F1 score of {\tt Xavier} models when classifying NVSS FR~Is and FR~IIs shares a common improvement. Typically, these models have their FR~II identification ability strengthened significantly. The precision of FR~II classification on NVSS test images reached 79$\%$, while the number when using PNG input was only 54$\%$. Nevertheless, when considering FIRST images, the models showed a balanced but 
relatively smaller recall, precision, and F1 score.

The {\tt Xavier} models trained with FIRST images also showed general improvement when identifying NVSS images. Though the issue of FR~I preference still exists, recall of FR~IIs classification increased by 39$\%$ compared to that using PNG inputs. This implies that by using JPEG images, the classifier achieved a higher sensitivity for identifying FR~IIs. 

The JPEG-based results also echo the transfer learning outcomes seen using PNG format. No matter which method was applied, models inheriting weights trained on FIRST images and then re-trained on NVSS images make a more accurate prediction. Typically, by applying Method 0, models work optimally for classifying both NVSS and FIRST images. When transfer learning models inherited weights from models trained on FIRST images, their performance is similar to that using PNG inputs. 

In spite of the similarities, we note that there are two other phenomena worth mentioning. The first is that the difference between randomly initialized models and those using Method 0 are reduced when using JPEG inputs. The accuracy difference between the two is less than 0.5$\%$, while the difference is larger than 1.5$\%$ using PNG formatted inputs. The second phenomenon is caused by applying Method A. The application of Method A no longer makes the best FIRST prediction if re-trained on NVSS images in JPEG format. If one adopted image input in JPEG when learning and testing, Methods 0 and B would become better options.

%

Why changing image format leads to overall model performance enhancement is not immediately obvious. To explain it, we consider the different input images from the perspective of their information content. We do this by evaluating the Shannon entropy of the pre-processed input images in FITS, PNG and JPEG formats.

Shannon entropy refers to the averaged self-information content of a dataset \citep{1949mtc..book.....S}. Self-information can be defined as the probability that a stochastic source of noise has produced the information in the dataset. Equation~\ref{eqn:IV} gives the mathematical definition of Shannon entropy, $S$,
\begin{ceqn}
\begin{align}
S = -\sum p_{k} \log p_{k}
\label{eqn:IV}
\end{align}
\end{ceqn}
where $p_{k}$ represents the normalized pixel values considered as probabilities. For this work, we adopt 2 as the logarithmic base when measuring Shannon entropy.

By definition, inputs with lower Shannon entropy have smaller variation . Also, since the image inputs in this work are normalized to the same pixel range $(0-255)$, an image with high Shannon entropy should have a weakly concentrated pixel value distribution. In other words, a model would find it easier to learn image pixel value gradients if the same image had higher Shannon entropy. 

We compared mean Shannon entropy between inputs in different data formats from different surveys. Table~\ref{tab:Shannon_Entropy_Comparison} provides a summary of these entropy measurements. In general, NVSS sample images have higher Shannon entropy than FIRST sample images. When we take Table~\ref{tab:transfer_leraning_accuracy} and Table~\ref{tab:transfer_leraning_accuracy_JPEG} into account, we find that most models re-trained with FIRST images tended to have a smaller standard deviation in accuracy.

When converting input images from the  CoNFIG catalog from FITS format to JPEG or PNG, we found that their Shannon entropy consistently dropped. Images in PNG format have the lowest mean Shannon entropy of all three formats. In the context of classification, models seem to make a more accurate predictions if the input data shares higher mean Shannon entropy. 
Since inputs in FITS format have the highest Shannon entropy, it is recommended that future networks use FITS inputs for both training and testing machine learning models.

Regardless of survey or catalog, FR~II inputs experiencing format conversion show higher fractional loss of entropy than FR~Is. 
However, we did not see a relationship between this loss and model performance. When we train and test model using data from the same survey, Recall, Precision, and F1 score differences between the two classes are less than 10$\%$. Such differences become more apparent when testing on a different survey, but the self-information imbalance between the two classes is not sufficient to explain the difference. When applying transfer learning, models continued to give comparative or more accurate FR~II and FR~I predictions on FIRST and NVSS images, respectively. 

Overall, how image formats affects model performance still requires further investigation. Whether Shannon entropy can be seen as an evaluation factor also needs examination in the future. 

\begin{table}
\begin{center}
\begin{tabular}{| c | c | c | c |}
\hline \hline
\multirow{1}{*}{\textbf{NVSS}} &   &    \\ 

\textbf{inputs} & \multicolumn{1}{ c |}{\textbf{FITS}}   & \textbf{PNG}  & \multicolumn{1}{ c |}{\textbf{JPEG}}  \\
\hline
CoNFIG~FR~I & 0.32$\pm$0.12   & 0.18$\pm$0.11 & 0.22$\pm$0.14    \\
CoNFIG~FR~II & 0.28$\pm$0.03   & 0.15$\pm$0.7 & 0.19$\pm$0.1   \\
FRICAT & 0.28$\pm$0.03   & 0.20$\pm$0.06 & 0.28$\pm$0.11    \\
\hline
\multirow{ 1}{*}{\textbf{FIRST}} &  &   \\ 

\textbf{inputs}& \multicolumn{1}{ c |}{\textbf{FITS}} & \textbf{PNG} & \multicolumn{1}{ c |}{\textbf{JPEG}}    \\
\hline
CoNFIG~FR~I & 0.25$\pm$0.16   & 0.15$\pm$0.09 & 0.19$\pm$0.11 \\
CoNFIG~FR~II & 0.14$\pm$0.09  & 0.06$\pm$0.05  & 0.07$\pm$0.07  \\
FRICAT & 0.09$\pm$0.05   & 0.07$\pm$0.03 & 0.19$\pm$0.09    \\
\hline \hline
\end{tabular}
\end{center}
\caption{A summary of Shannon entropy measurement for image inputs in different formats. Shannon entropy for inputs in FITS format have experienced image pre-processing.}
\label{tab:Shannon_Entropy_Comparison}
\end{table}

\section{The application of transfer learning to future radio surveys}
\label{sec5_0}

\begin{figure}
\setlength{\unitlength}{1cm}
\begin{center}
\includegraphics[width=0.5\textwidth]{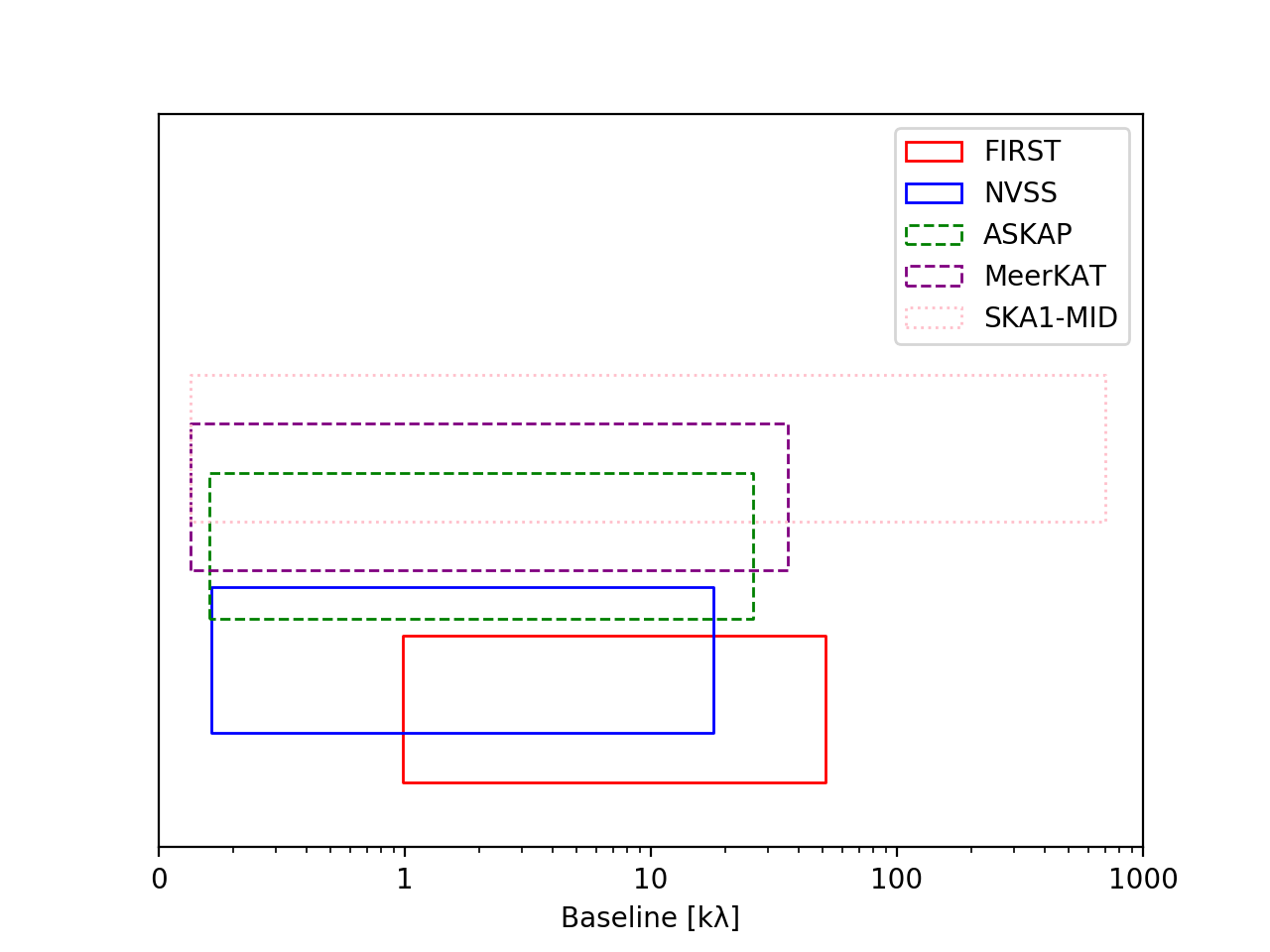}
\end{center}
\caption{A summary of spatial scales for several radio telescopes/surveys in units of kilo-lambda (k$\lambda$). Solid: finished radio surveys. Dashed: radio telescopes (almost) finish construction. Dotted: telescope would be built in the future. Spatial scales shown in the diagram are converted from telescope baselines in units of {\tt km}. The frequency adopted when doing the conversion is 1.4 GHz for FIRST, NVSS, MeerKAT and SKA1-MID. We adopted 1.3 GHz for ASKAP specifically for its EMU survey \citep{2011PASA...28..215N}. FIRST was observed using the VLA B-configuration of the VLA \citep{1995ApJ...450..559B}, while NVSS adopted the more compact D and DnC configurations of the same array \citep{1998AJ....115.1693C}. ASKAP have minimum and maximum baseline of 37 m and 6 km, respectively \citep{2008ExA....22..151J,2015MNRAS.452.2680S}. Baselines of MeerKAT ranges from 29 m to 7 km \citep{2016mks..confE...1J}. Finally, SKA1-MID is expected to have 150 km maximum baseline. The shortest baseline of SKA1-MID here is the same as MeerKAT, as MeerKAT will finally become a part of SKA1-MID core \citep{Braun2015}. }
\label{fig:baseline_comparison}
\end{figure}


Traditionally, radio galaxy classification has been done by visual inspection, sometimes facilitated by measurement of host-hotspot relative positions. Such a method was practical due to the modest sample size of archival catalogs. However, soon next-generation radio catalogs such as that from the EMU survey \citep{2011PASA...28..215N}, will discover millions of radio sources waiting for visual inspection. 

In order to overcome the difficulties of classifying these sources by eye, recent studies have focused on developing machine-learning based automated methods to classify radio source morphologies based on specific radio surveys. In this paper, we have introduced the next step in the use of these methods and explored the possibility to boost model performance by applying transfer learning.

 Our approaches achieved over 90.1$\%$ and 83.9$\%$ in terms of classification accuracy when testing on FIRST and NVSS images, comparable with other recent state-of-the-art results. Depending on the transfer learning method used, we have demonstrated that transfer learning models can result in even higher model  accuracies or save training time by up to 79$\%$. 

A key result from this work is that inheriting model weights pre-trained on higher resolution survey data, e.g. FIRST, can boost model performance when re-training with lower resolution survey images, e.g. NVSS. However, we found that the reverse situation, whereby weights inherited from models trained on lower resolution data are re-trained on higher resolution data, is detrimental to model performance. This is of particular relevance for future radio surveys, where machine learning weights inherited from models trained on archival data may be used to initiate classifiers for previously unseen data. Figure~\ref{fig:baseline_comparison} summarizes the baseline ranges of the  NVSS and FIRST surveys, along with the ranges of ASKAP, MeerKAT, and SKA1-MID. These three telescopes are capable of making observations at 1.3-1.4\,GHz, similar to FIRST and NVSS which were made at 1.4\,GHz. It can be seen that there are considerable spatial scale overlap between MeerKAT, ASKAP and the surveys considered in this work. 

The higher resolution of the FIRST survey, relative to both MeerKAT and ASKAP as well as NVSS, suggests the potential for successful transfer learning approaches to machine learning classification of the survey data from these next-generation telescopes; however, the significantly improved resolution of SKA1-MID in comparison suggests that further investigations must be made before the advantages of transfer learning can be used there. 


\section*{Acknowledgements}

The authors are very grateful for discussions from the machine learning group at Jodrell Bank Centre for Astrophysics, JBCA. This research was supported by JBCA, University of Manchester.








\appendix

\section{Some extra material}

If you want to present additional material which would interrupt the flow of the main paper,
it can be placed in an Appendix which appears after the list of references.




\bsp	
\label{lastpage}
\end{document}